\documentclass[prd,aps,amsmath,amssymb,letter]{revtex4}

\usepackage{graphicx}% Include figure files
\usepackage{dcolumn}% Align table columns on decimal point
\usepackage{bm}% bold math
\usepackage{epsfig,latexsym}

\newcommand{\be}{\begin{eqnarray}}
\newcommand{\ee}{\end{eqnarray}}
\newcommand{\ba}{\begin{array}}
\newcommand{\ea}{\end{array}}

\begin{document}
\title{Quarkonia Correlators Above Deconfinement}
\author{\'{A}gnes {\sc M\'{o}csy}}
\email{mocsy@bnl.gov}  \affiliation{RIKEN-BNL Research Center,
Brookhaven National Laboratory, Upton, New York 11973, USA }
\affiliation{FIAS \& Theoretical Physics Institute,
J.W. Goethe-Universit\"at, % Max von Laue Str 1,
60438 Frankfurt am
Main, Germany}
\author{P\'eter {\sc Petreczky}}
\email{petreczk@quark.phy.bnl.gov} \affiliation{RIKEN-BNL Research
Center \& Department of Physics, Brookhaven National Laboratory,
Upton, New York 11973, USA}
\date{\today}

\begin{abstract}
We study the quarkonia correlators above deconfinement using the
potential model with screened temperature-dependent potentials. We
find that while the qualitative features of the spectral
functions, such as the survival of the 1S state, can be reproduced
by potential models, the temperature dependence of the correlators
disagree with the recent lattice data.
\end{abstract}

\maketitle

%%%%%%%%%%%%%%%%%%%%%%
\section{Introduction}
\label{intro}

Current and past experiments at BNL and at CERN are and have been
colliding heavy ions at relativistic velocities. One of the major
goals of these experiments is to produce a deconfined state of
matter, known as quark-gluon plasma. In this deconfined phase the
dominant degrees of freedom are the different flavors of quarks
and gluons which are not bound inside hadrons anymore. Unlike for
light quarks, due to their smaller size, bound states of heavy
quarks could survive inside the plasma to temperatures higher than
that of deconfinement, $T_c$. In the 1980's, however, it was
predicted \cite{Matsui:1986dk} that $c\bar{c}$ bound states would
disappear already at temperatures close to $T_c$. The idea of
Matsui and Satz, based on non-relativistic arguments, was that
color screening in the plasma would prevent the strong binding of
quarkonia. Therefore, the dissolution of heavy quark bound states,
and thus the suppression of the $J/\psi$ peak in the dilepton
spectra, could signal deconfinement.

It was recognized in the 1970's \cite{Eichten:1979ms} that what is
now known as the Cornell potential (a Coulomb plus a linear part)
provides a very good description of the quarkonia spectra at zero
temperature. Much later it was understood that in non-relativistic
systems, such as quarkonium, the hierarchy of energy scales $m \gg
m v \gg m v^2$ ($m$ and $v$ being the heavy quark mass and
velocity), allows the construction of a sequence of effective
theories : nonrelativistic QCD (NRQCD) and potential NRQCD
(pNRQCD) \cite{Bali:2000gf,yellow}. The potential model appears as
the leading order approximation of pNRQCD
\cite{Bali:2000gf,yellow}, and thus can be derived in QCD from
first principles.

The essence of the potential model calculations in the context of
deconfinement is to use a finite temperature extension of the zero
temperature Cornell potential to understand the modifications of
the quarkonia properties at finite temperature
\cite{Karsch:1987pv,{Digal:2001ue}}. It is not a priori clear
though, whether the medium modifications of the quarkonia
properties can be understood in terms of a temperature dependent
potential. Therefore, a new way of looking at this problem has
been developed. This new way is based on the evaluation on the
lattice of the correlators and spectral functions of the heavy
quark states \cite{Umeda:2002vr,Asakawa:2003re,{Datta:2003ww}}.
But why are correlators of heavy quarks of interest? As we said
above, hadronic bound states are expected to dissolve at high
temperatures. With increasing temperature such resonances become
broader and thus very unstable and gradually it becomes
meaningless to talk about them as being resonances. Accordingly,
at and/or above $T_c$ they stop being the correct degrees of
freedom. Correlation functions of hadronic currents, on the other
hand, are meaningful above and below the transition. These can
thus be used in a rather unambiguous manner to extract and follow
the modification of the properties of quarkonia in a hot medium.

The numerical analysis of the quarkonia correlators and spectral
functions carried out on the lattice for quenched QCD provided
unexpected, yet interesting results
\cite{Umeda:2002vr,Asakawa:2003re,Datta:2003ww}. The results
suggest the following: The ground state charmonia, 1S $J/\psi$ and
$\eta_c$, survive well above $T_c$, at least up to $1.6T_c$. Not
only do these states not melt at or close to $T_c$ as was
expected, but lattice found little change in their properties when
crossing the transition temperature. In particular, the masses of
the 1S states show almost no thermal shift
\cite{Umeda:2002vr,Asakawa:2003re,Datta:2003ww}. Also, although
the temperature dependence of the ground state quarkonium
correlators was found to be small, a small difference between the
behavior of the $J/\psi$ and the $\eta_c$ correlators has been
identified. This difference was also not a priori expected.
Furthermore, the lattice results indicate \cite{Datta:2003ww} that
properties of the 1P states, $\chi_{c0}$ and $\chi_{c1}$, are
significantly modified above the transition temperature, and these
states have dissolved already at $1.1T_c$. The recently available
first results for the bottomonia states \cite{Petrov:2005ej}
suggests, that the $\eta_b$ is not modified until about $2T_c$,
but the $\chi_{b0}$ shows dramatic changes already at $1.15T_c$.

These features from the lattice calculations are in sharp contrast
with earlier, potential model studies of quarkonium properties at
finite temperature, which predicted the dissolution of charmonia
states for $T \le 1.1T_c$ \cite{Karsch:1987pv,Digal:2001ue}. More
recent studies try to resolve this apparent contradiction by
choosing a potential which can bind the heavy quark anti-quark
pair at higher temperatures. In \cite{Shuryak:2003ty} the picture
of strongly coupled Coulomb bound states is discussed. In
\cite{Wong:2004zr,Alberico:2005xw,{Mannarelli:2005pa}} the
internal energy of a static quark anti-quark pair calculated on
the lattice at finite temperature is used as the potential. It is
found that the $J/\psi$ could be bound at temperatures as high as
$2T_c$, in agreement with the lattice studies of the spectral
functions. The quarkonium properties, however, are significantly
modified in all of these calculations.

In connection with the potential models two questions should be
addressed: First, can the medium modifications of quarkonia
properties, and eventually their dissolution be understood in
terms of a temperature dependent potential? Second, what kind of
screened potential should be used in the Schr\"odinger equation,
that describes well the bound states at finite temperature? At
present neither of these questions have been answered, since it is
not clear whether the sequence of effective theories leading to
the potential model can be derived also at non-zero temperature,
where the additional scales of order $T$, $gT$ and $g^2 T$ are
present. Therefore, in this paper we follow a more
phenomenological approach, extending our results presented in
\cite{Mocsy:2004bv}. We address the first question by constructing
the heavy quarkonia correlators within the potential model and
comparing them to the available lattice data. Our way of
addressing the second question is to consider two different
screened potentials and show that the qualitative conclusions do
not depend on the specific choice of the potential. Our study of
quarkonia correlators in Euclidean time is also motivated by the
fact that although lattice calculations give very precise results
about the temperature dependence of the quarkonia correlators (see
e.g. \cite{Datta:2003ww}), it is quite difficult to reconstruct
the spectral functions from the correlators at finite temperature.

The paper is organized as follows: Section \ref{model} presents
the model we use to construct the spectral function and analyze
the correlators. In Section \ref{res-prop} our results for the
temperature dependence of the quarkonia properties are presented.
Section \ref{res-corrI} is devoted to the study of charmonium and
bottomonium correlators in the scalar and pseudoscalar channels.
In Section \ref{res-corrII} we discuss the vector correlator which
also carries information about diffusion properties. We summarize
the results and present our conclusions in Section \ref{conclude}.

%%%%%%%%%%%%%%%%%%%%%%%%%%%%%%%%%%%%%%%%%%%%%%%%%%%%%%%
\section{Euclidean correlators and the potential model}
\label{model}

Here we investigate the quarkonium correlators in Euclidean time
at finite temperature, as these correlators are directly
calculable in lattice QCD. To make contact with the available
lattice data obtained in the quenched approximation,
i.e.~neglecting the effects of light dynamical quarks,  we
consider QCD with only one heavy quark flavor. Furthermore, we
only consider the case of zero spatial momentum.

The correlation function for a particular mesonic channel $H$ is
defined as
\be
G_H(\tau,T)=\langle  j_H(\tau)j_H^\dagger(0)\rangle\, .
\label{corrdef}\ee Here $j_H = \bar{q} \Gamma_H q$, and $\Gamma_H
= 1, \gamma_\mu, \gamma_5, \gamma_\mu\gamma_5$ corresponds
respectively to the scalar, vector, pseudoscalar and axial vector
channels. At zero temperature these channels correspond to the
different quarkonium states shown in Table \ref{tab:channels}.
\begin{table}[h]
\renewcommand{\arraystretch}{1.2}
\begin{center}
\begin{minipage}{8.3cm} \tabcolsep 5pt
\begin{ruledtabular} \caption{Quarkonia channels considered.}
\begin{tabular}{||cccccc||}
 &$j_H$&$^{2S+1}L_J$&$J^{PC}$&$q=c$&$q=b$\\ \hline scalar&$\bar{q}q$&
$^3P_0$&$0^{++}$&$\chi_{c0}$&$\chi_{b0}$\\
pseudoscalar&$\bar{q}\gamma_5q$&
$^1S_0$&$0^{-+}$&$\eta_c$&$\eta_b$\\ vector&$\bar{q}\gamma_\mu q$&
$^3S_1$&$1^{--}$&$J/\psi$&$\Upsilon$\\ axial
vector&$\bar{q}\gamma_\mu\gamma_5q$&
$^3P_1$&$1^{++}$&$\chi_{c1}$&$\chi_{b1}$\\
\end{tabular}
\label{tab:channels}
\end{ruledtabular}
\end{minipage}
\end{center}
\end{table}
Using the Euclidean Hamiltonian (transfer matrix) the spectral
decomposition of the correlators defined in Eq.~(\ref{corrdef}) at
zero temperature can be written as \cite{montvay}
\be
G_H(\tau,T=0)=\sum_n|\langle  0|j_H|n\rangle|^2 e^{-E_n \tau} \, ,
\label{expv}\ee where the $E_n$ are the eigenstates of the
Hamiltonian. For the few lowest lying states $E_n$ corresponds to
the quarkonia masses, while the matrix elements $|\langle  0|j_H |
n\rangle|^2=F_i^2$ give the decay constants. Furthermore, the
Euclidean correlator is an analytic continuation of the real time
correlator $D^{>}(t)$, $G_H(\tau,T)=D^{>}_H(-i\tau,T)$, and the
spectral function is
\be
\sigma(\omega,T)=\frac{D^{>}_H(\omega,T)-D^{<}_H(\omega,T)}{2
\pi}= \frac{1}{\pi} {\rm Im} D^{R}_H(\omega,T)\, . \ee Here
$D^{R}_H(\omega,T)$, $D^{>}(\omega,T)$ and $D^{<}(\omega,T)$ are
the Fourier transforms of the real time correlators
\begin{eqnarray}
& D^R(t,T)=\langle [j(t),j(0)]\rangle \, ,\\ & D^{>}(t,T)=\langle
j(t) j(0)\rangle,~~D^{<}(t,T)=\langle j(0) j(t)\rangle\, ,
\end{eqnarray}
and $\langle ...\rangle$ denotes the expectation value at finite
temperature $T$. Using the above, the following spectral
representation for $G_H(\tau,T)$ can be derived (see Appendix
\ref{app-spectral})
\be
G(\tau,T)=\int d\omega \sigma(\omega,T) K(\tau,\omega,T) \, ,
\label{f}\ee where $K$ is the integration kernel
\be
K(\tau,\omega,T)=\frac{\cosh{\left[\omega\left(\tau-\frac{1}{2T}
\right)\right]}} {\sinh{\left[\frac{\omega}{2T}\right]}} \,
.\label{kernel}\ee

Recent lattice QCD calculations utilize the form (\ref{f}) for the
correlator to extract spectral functions. Results obtained with
the Maximum Entropy Method are promising, but still somewhat
controversial, and not fully reliable. We choose to directly
analyze the correlators that are reliably calculated on the
lattice. So as an input to (\ref{f}) we need to specify the
spectral function at finite temperature. We do this by following
the form proposed in \cite{Shuryak:kg} for the zero temperature
spectral function
\be
\sigma(\omega)=\sum_i 2 M_i F_i^2 \delta\left(\omega^2
-M_i^2\right) + \frac{3}{8 \pi^2} \omega^2
\theta\left(\omega-s_0\right)f(\omega,s_0)\, . \label{spf} \ee The
first term contains the pole contributions from bound states
(resonances), while the second term is the perturbative continuum
above some threshold, denoted here by $s_0$. While we know the
asymptotic behavior of the spectral function from perturbation
theory, no reliable information is available for the explicit form
of the threshold function, $f(\omega,s_0)$. Therefore, in this
work we consider two simple choices for it: First, we take the
most simple form of $f(\omega,s_0)=a_H$, valid for free massless
quarks, resulting in a sharp threshold. Then, we consider the form
motivated by leading order perturbative calculations with massive
quarks,
\be
f(\omega)=\left(a_H + b_H \frac{s_0}{\omega^2} \right)
\sqrt{1-\frac{s_0^2}{\omega^2}}\, ,\label{smooth} \ee giving a
smooth threshold. In leading order perturbation theory
$s_0=2m_{b,c}$. The coefficients ($a_H,b_H$) were calculated at
leading order in \cite{Karsch:2000gi}, and are (-1,1), (1,0),
(2,1), and (-2,3), for the scalar, pseudoscalar, vector, and axial
vector channels, respectively \footnote{The meson correlators in
\cite{Karsch:2000gi} differ by a factor of two from our
definition.}. Comparing the results obtained with these two forms
of the threshold function gives us an estimate on the
uncertainties resulting from the simplified form of the spectral
function in equation (\ref{spf}).

In real QCD with three light quarks it is natural to identify the
continuum threshold $s_0$ with the open charm or beauty threshold.
For the case of one heavy quark only, which is considered here,
the value of $s_0$ is somewhat arbitrary. Above some energy
though, the spacing between the different states is so small that
they eventually form a continuum. Parton-hadron duality then
requires, that the area under the spectral function above this
energy range is the same as the area under the free perturbative
spectral function, justifying the form given above. This also
motivates our choice of threshold as the energy above which no
individual resonances are observed experimentally. The parameters
of the zero temperature spectral function are given in Table
\ref{tab:zerotemp}.
\begin{table}[h]
\renewcommand{\arraystretch}{1.2}
\begin{center}
\begin{minipage}{10cm} \tabcolsep 5pt
\begin{ruledtabular} \caption{Parameters at $T=0$.}
\begin{tabular}{||c|c|c|c|c|c||}
$\alpha$&$\sigma$&$m_c$&$m_b$&$s_{0c}$&$s_{0b}$\\ \hline
0.471&0.192 GeV$^2$ &1.32 GeV&4.746 GeV&4.5 GeV&11 GeV\\
\end{tabular}
\label{tab:zerotemp}
\end{ruledtabular}
\end{minipage}
\end{center}
\end{table}

The remaining parameters of the spectral function (\ref{spf}) can
be calculated using a potential model. The bound state masses are
given by $M_i = 2m + E_i$, where $E_i$ are the binding energies,
and $m$ is the constituent mass of either the charm or the bottom
quark. In order to calculate the decay constants $F_i$, we first
write the relativistic quark fields in terms of their
non-relativistic components. This can be done using the
Foldy-Wouthysen-Tani transformation
\be
q=\exp\left(\frac{{\bm \gamma \cdot \bm D}}{2 m}\right) \left(
\begin{array}{c} \psi \\ \chi
\end{array}
\right)\, , \ee with $\bm D$ being the spatial covariant
derivative. At leading order in the coupling and inverse mass, the
decay constant can be related to the wave function at the origin
for the S states, i.e.~the pseudoscalar and vector channel
\cite{Bodwin:1994jh},
\be
F_{PS}^2 = \frac{N_c}{2\pi}|R_n(0)|^2 \qquad \mbox{and}\qquad
F_{V}^2 = \frac{3N_c}{2\pi}|R_{n}(0)|^2\, , \label{ps-v}\ee and to
the derivative of the wave function in the origin for P states,
i.e.~the scalar and axial-vector channels \cite{Bodwin:1994jh},
\be
F_{S}^2 = -\frac{9N_c}{2\pi m^2}|R'_n(0)|^2 \qquad
\mbox{and}\qquad F_{A}^2 = -\frac{9N_c}{\pi m^2}|R'_n(0)|^2\, ,
\label{s-a}\ee where $n=1,2,...$. For the number of colors we
consider $N_c=3$. The negative signs in Eq.~(\ref{s-a}) are the
consequences of our definitions of the Euclidean Dirac matrices.
For the derivation of these relations see Appendix
\ref{app-bound}. To obtain the binding energies and the wave
functions we solve the Schr\"odinger equation
\be
\left(-\frac{1}{m}\frac{d^2}{dr^2} + \frac{l(l+1)}{mr^2} + V(r) -
E \right)u(r)=0\, , \qquad R(r)=\frac{u(r)}{r}\, , \label{schr}
\ee with the Cornell potential
\be
V(r)=-\frac{\alpha}{r}+\sigma r\, . \label{cornellpot}\ee Here
$\alpha$ is the coupling and $\sigma$ is the string tension. Their
values, as well as the quark masses, were obtained in
\cite{Jacobs:1986gv} by fitting the zero temperature quarkonium
spectrum. The parameters of the zero temperature analysis are
summarized in Table \ref{tab:zerotemp}. Note that the angular
momentum $l$ in (\ref{schr}) distinguishes between the different
quarkonium states shown in Table \ref{tab:channels}.

To model the binding and propagation of a heavy quark and
antiquark in the deconfined phase, we assume that they interact
via a temperature-dependent screened potential, and propagate
freely above some threshold. Then, our finite temperature model
spectral function has the form given in (\ref{spf}) with a now
temperature-dependent decay constant, quarkonium mass, and
threshold:
\be
\sigma(\omega,T)=\sum_i 2 M_i(T) F_i(T)^2 \delta\left(\omega^2
-M_i(T)^2\right) + m_0 \omega^2
\theta\left(\omega-s_0(T)\right)f(\omega, s_0(T)) +
\chi_s(T)q(T)\omega\delta(\omega) \, . \label{spft} \ee The last
term in the spectral function is important at low frequencies and
contributes at nonzero temperatures. It is present only in the
vector channel, and is due to charge fluctuations and diffusion.
The function $q(T)=0$ for the scalar and the pseudoscalar states,
and $q(T)=(3T/m-1)$, for the vector channel. Since we consider the
case of zero spatial momentum, there is no distinction between the
transverse and the longitudinal components in the vector channel.
Therefore, we sum over all four Lorentz components,
$\sigma_V=\sum_{\mu\nu}\sigma_{\mu\nu}$. At nonzero spatial
momentum, for $\omega^2-k^2<0$, there would also be an additional
contribution to the spectral function \cite{Karsch:2003wy}. The
charge susceptibility $\chi_s$ in the nonrelativistic
approximation has the following form
\be
\chi_s(T) = 4N_c\frac{1}{(2\pi)^{3/2}} m^{3/2} T^{1/2} e^{-m/T}\,
. \ee The derivation of this expression, together with the full
evaluation of the vector correlator at one-loop level is provided
in the Appendix \ref{transport}. The relevance of this
contribution in the spectral function for the possibility of
obtaining transport coefficients from the lattice has been
recently discussed in \cite{Petreczky:2005nh}.

\begin{figure}[h]
\begin{minipage}[h]{5.5cm}
\epsfig{file=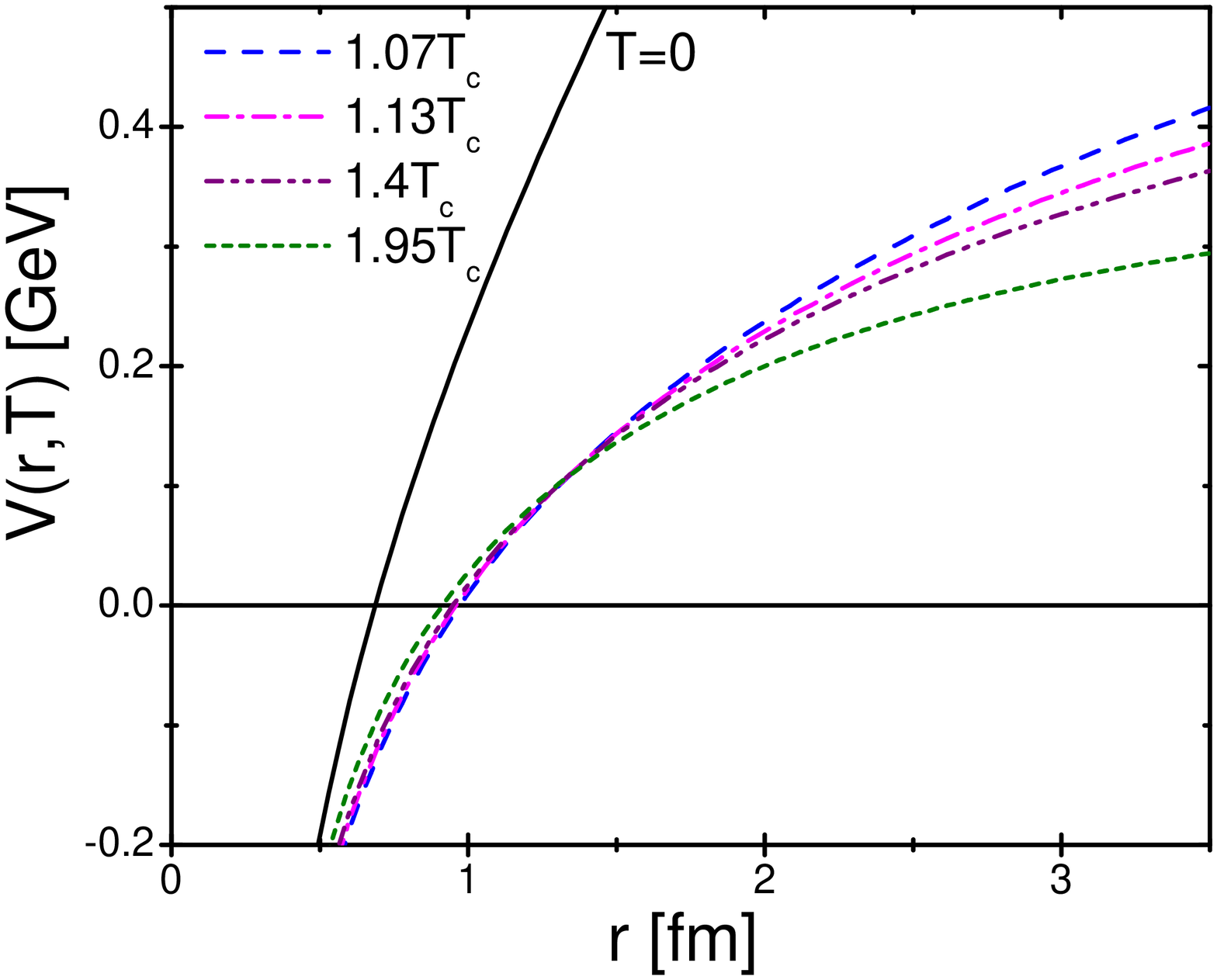,height=53mm}
\end{minipage}
\hspace*{2cm}
\begin{minipage}[h]{5.5cm}
\epsfig{file=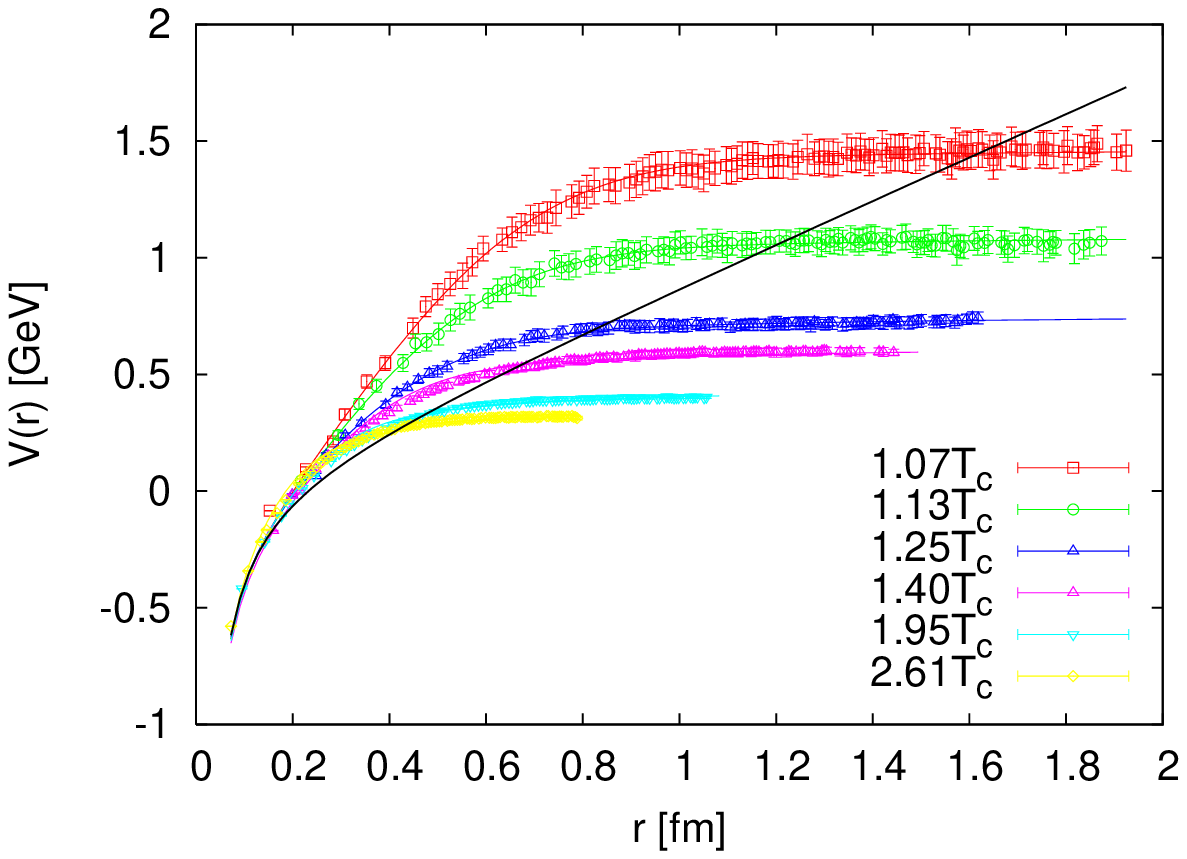,height=53mm}
\end{minipage}
\caption{Cornell potential at $T=0$ and screened Cornell potential
for different $T>T_c$ (left panel); Lattice data on internal
energy from \cite{Kaczmarek:2003dp} together with the fit to the
data (right panel).} \label{fig:pot}
\end{figure}

Just like at zero temperature, we calculate the binding energy and
the wave function of the quarkonium states at finite temperature
using the Schr\"odinger equation (\ref{schr}). For this we need to
specify a temperature-dependent screened quark-antiquark
potential. While the zero temperature potential can be calculated
on the lattice, at nonzero temperature it is not clear how to
define this quantity (for a discussion of this see
\cite{Petreczky:2005bd}). Some ideas on how to generalize the
notion of static potential to temperatures above deconfinement
were presented in \cite{Simonov:2005jj}. In the current analysis
we use two different potentials: The screened Cornell potential
above $T_c$, first considered in \cite{Karsch:1987pv}
\be
V(r,T)=-\frac{\alpha}{r}e^{-\mu(T) r}+\frac{\sigma}{\mu(T)}
(1-e^{-\mu(T) r}) \, . \label{pot}\ee The coupling and the string
tension are specified in Table \ref{tab:zerotemp}. We choose the
following Ansatz for the temperature dependence of the screening
mass $\mu=[0.24+0.31 \cdot(T/T_c-1)]~$GeV with $T_c=0.270~$GeV.
This choice is motivated by the fact that we want to reproduce the
basic observation from the lattice, namely that the 1S state
exists up to $1.6T_c$, while the 1P state melts at $1.1T_c$. The
potential (\ref{pot}) is shown in the left panel of Figure
\ref{fig:pot} for different temperatures above deconfinement.

In \cite{Shuryak:2003ty,Wong:2004zr,Alberico:2005xw} the change in
the internal energy induced by a static quark-antiquark pair,
first calculated in \cite{Kaczmarek:2003dp}, is used as the
potential. This is a debatable choice of potential, since near the
transition temperature the internal energy shows a very large
increase
\cite{{Petreczky:2005bd},Petreczky:2004pz,Petreczky:2004xs}.
However, for temperatures $T>1.07T_c$ we will also use the
internal energy calculated on the lattice \cite{Kaczmarek:2003dp}
as the potential. We parameterize this as
\be
V(r,T) &=& -\frac{\alpha}{r}e^{-\mu(T) r^2} + \sigma(T) r
e^{-\mu(T) r^2} + C(T)\left(1-e^{-\mu(T) r^2}\right)\,
.\label{latpot}\ee This parameterization is shown against the
actual lattice data in the right panel of Figure \ref{fig:pot}.
The temperature dependence of the parameters of (\ref{latpot}) is
shown in Tables \ref{tab:latc} and \ref{tab:latb} in Appendix
\ref{app:tables}.
\begin{figure}
\epsfig{file=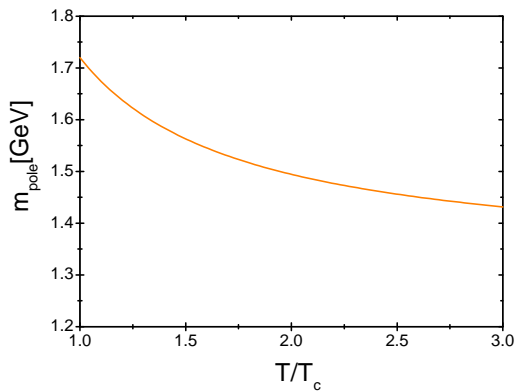,height=53mm} \caption{temperature
dependence of the quark pole mass.} \label{fig:polemass}
\end{figure}

In the presence of screening the potential has a finite value at
infinite separation $V_{\infty}(T)=\mbox{lim}_{r\rightarrow\infty}
V(r,T)$ (see Fig.~\ref{fig:pot}). The $V_{\infty}(T)$ is the extra
thermal energy of an isolated quark. The minimum energy above
which the quark-antiquark pair can freely propagate is
$2m+V_{\infty}$. In what follows, we assume that this minimum
energy defines the continuum threshold,
i.e.~$s_0(T)=2m+V_{\infty}$. Independently of the detailed form of
the potential, $V_{\infty}(T)$ is decreasing, and thus $s_0$ will
also decrease with temperature. This effect manifests in the
temperature dependence of the correlators, as we will discuss in
the following Section. Since at leading order the threshold of the
continuum is $2m$, we may think of $s_0(T)/2$ as the temperature
dependent pole mass, shown in Fig.~\ref{fig:polemass}. The
decrease of $V_{\infty}(T)$ near $T_c$ can thus be considered as a
decrease of the pole mass. Such a decrease of the pole mass was
observed in lattice calculations \cite{Petreczky:2001yp}, when
calculating quark and gluon propagators in the Coulomb gauge. As
the temperature increases the position of the peak will be close
to the threshold, causing the phenomenon of threshold enhancement.
This has recently been discussed for light quarks in
\cite{Casalderrey-Solana:2004dc}. We will not consider this in the
present paper. While we assume that above the threshold quarks and
antiquarks propagate with the temperature dependent effective mass
defined above, quarks inside a singlet bound state will not feel
the effect of the medium, and thus will have the vacuum mass.
Therefore in the Schr\"odinger equation we use the zero
temperature masses of the c and b quarks. Furthermore, above
deconfinement quarkonium can also dissociate via its interaction
with gluons \cite{Kharzeev:1994pz}. This effect leads to a finite
thermal width, which we will also neglect in the current analysis.

Finally, in order to make a direct comparison with the lattice
results we normalize the correlation function by the so-called
reconstructed correlators \cite{Datta:2003ww,Petrov:2005ej}. This
is done to eliminate the trivial temperature dependence from the
kernel (\ref{kernel}) in the correlator (\ref{f}). The
reconstructed correlators are calculated using the spectral
functions at a temperature below the critical one, here at $T=0$:
\be
G_{recon}(\tau,T)=\int d\omega \sigma(\omega,T=0) K(\tau,\omega,T)
\, .\ee The ratio $G/G_{recon}$ can therefore indicate
modifications to the spectral function above $T_c$. Any difference
of this ratio from one is an indication of the medium effects.

%%%%%%%%%%%%%%%%%%%%%%%%%%%%%%%%%%%%%%%%%%%%%%%%%%
\section{Quarkonia properties above deconfinement}
\label{res-prop}

We first present the properties of the charmonia and bottomonia
states in the deconfined medium. These results are obtained by
solving the Schr\"odinger equation (\ref{schr}) with the screened
Cornell potential (\ref{pot}).
\begin{figure}[h]
\begin{minipage}[h]{5.5cm}
\epsfig{file=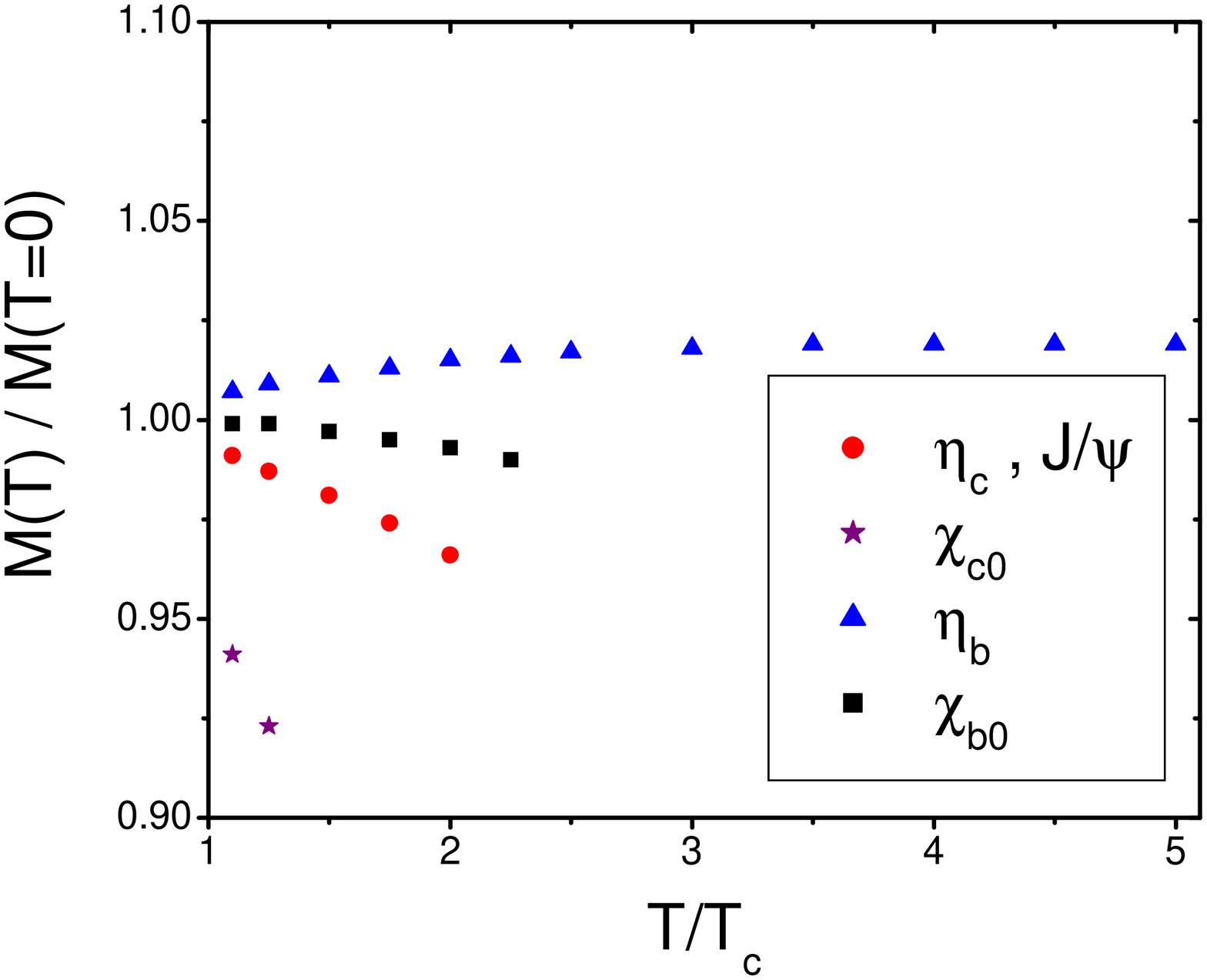,height=53mm}
\end{minipage}
\hspace*{2cm}
\begin{minipage}[h]{5.5cm}
\epsfig{file=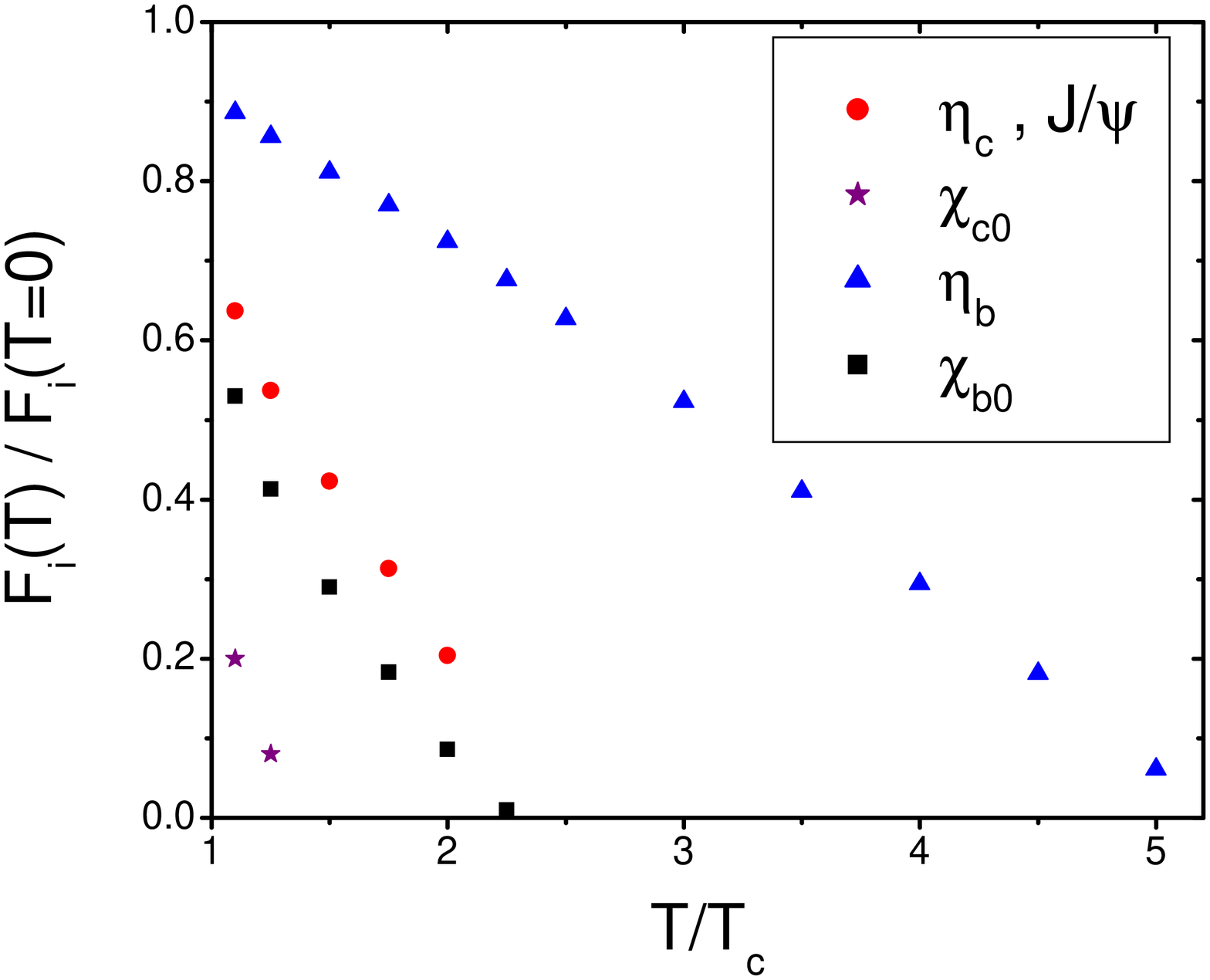,height=53mm}
\end{minipage}
\caption{Temperature dependence of quarkonia masses (left panel)
and amplitudes (right panel) normalized to their corresponding
zero temperature values.} \label{fig:mass-amp}
\end{figure}

The temperature dependencies of the masses and the amplitudes of
the different quarkonia states, normalized to their corresponding
zero temperature values, are shown in Fig. \ref{fig:mass-amp}. The
left panel of Fig. \ref{fig:mass-amp} shows that the bound state
masses do not change substantially with temperature. An exception
is the mass of the scalar charmonium $\chi_{c0}$, which shows a
significant decrease just above the transition temperature. The
right panel of Fig. \ref{fig:mass-amp} shows the amplitudes, as
obtained from the wave function, or its derivative at the origin.
Contrary to the masses, these show a strong drop with increasing
temperature for all the states considered. Since we neglect
effects that could arise from the hyperfine splitting, the
properties of the 1S scalar $\eta_c$ and pseudoscalar $J/\psi$
states are identical. While the small shift in the quarkonia
masses above the deconfinement temperature is consistent with
lattice data, the decrease in the amplitudes is neither confirmed,
nor ruled out by existing lattice data.
\begin{figure}[htbp]
\begin{center}
\resizebox{0.36\textwidth}{!}{%
  \includegraphics{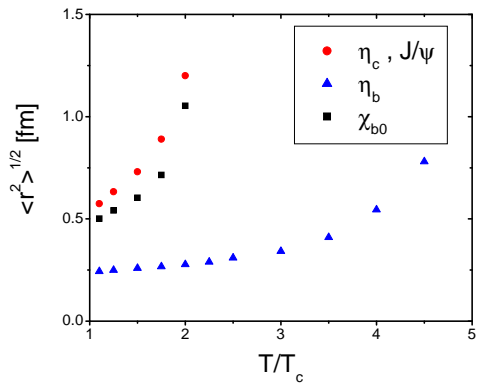}
}\caption{Temperature dependence of quarkonia radii.}
\label{fig:radius}
\end{center}
\end{figure}

In Fig.~\ref{fig:radius} the temperature dependence of the radii
is displayed. The radii of the scalar and pseudoscalar bottomonia
states begin to increase at much higher temperatures than their
corresponding charmonia states. This is expected, because
bottomonia is much smaller in size, and therefore survives to much
higher temperatures than charmonia. As the temperature increases
and the screening radius decreases the effective size of a bound
state becomes larger and larger, as seen in Fig.~\ref{fig:radius}.
When the size of the bound state is several times larger than the
screening radius, thermal fluctuations can dissociate it. This
means that the spectral functions no longer correspond to well
defined bound states, but rather to some very broad structures.
Here we consider a state to be melted, when its radius becomes
greater than 1 fm. For this reason we do not show the radius of
the $\chi_{c0}$, since it is significantly increased already at
$T_c$, suggesting the disappearance of this state at this
temperature. Note, that the $\chi_{b0}$ is similar in size to the
$\eta_c$ and the $J/\psi$, and shows significant increase of its
radius at about the same temperature.

As mentioned in the previous Sections, several recent studies used
the internal energy of a static quark-antiquark pair determined on
the lattice as the potential in the Schr\"odinger equation to
determine the properties of the quarkonia states at finite
temperature. In the present study we also consider this
possibility. The right panel in Figure \ref{fig:pot} shows that
close to $T_c$ the internal energy is significantly larger than
its zero temperature value. Therefore, when used in the
Schr\"odinger equation, this potential yields a large increase of
the quarkonia masses and amplitudes near $T_c$. The shift in the
quarkonia masses is of the order of several hundred MeV, and thus
disfavored by the lattice calculations of the quarkonia spectral
functions.

The numerical values for the quarkonia properties, together with
the temperature dependence of the parameters of the potential are
given in the Appendix \ref{app:tables}. Tables \ref{tab:c} and
\ref{tab:b} refer to the screened Cornell potential, and Tables
\ref{tab:latc} and \ref{tab:latb} to the potential identified with
the internal energy of a static quark-antiquark pair calculated on
the lattice.

%%%%%%%%%%%%%%%%%%%%%%%%%%%%%%%%%%%%%%%%%%%%%%%%%%%%%%%%%%%%%%&%%%%%%%%
\section{Numerical results for the Scalar and Pseudoscalar correlators}
\label{res-corrI}

In this Section we discuss the results obtained for the
temperature dependence of the correlators in the scalar and
pseudoscalar quarkonia channels. We compare these to existing
lattice results.

%%%%%%%%%%%%%%%%%%%%%%%%%%%%%%%%%%%%%%%%%%%%%%%%%%%%%%%%
\subsection{Results with the Screened Cornell Potential}
\label{cornell}

\begin{figure}[htbp]
\begin{minipage}[htbp]{5.5cm}
\epsfig{file=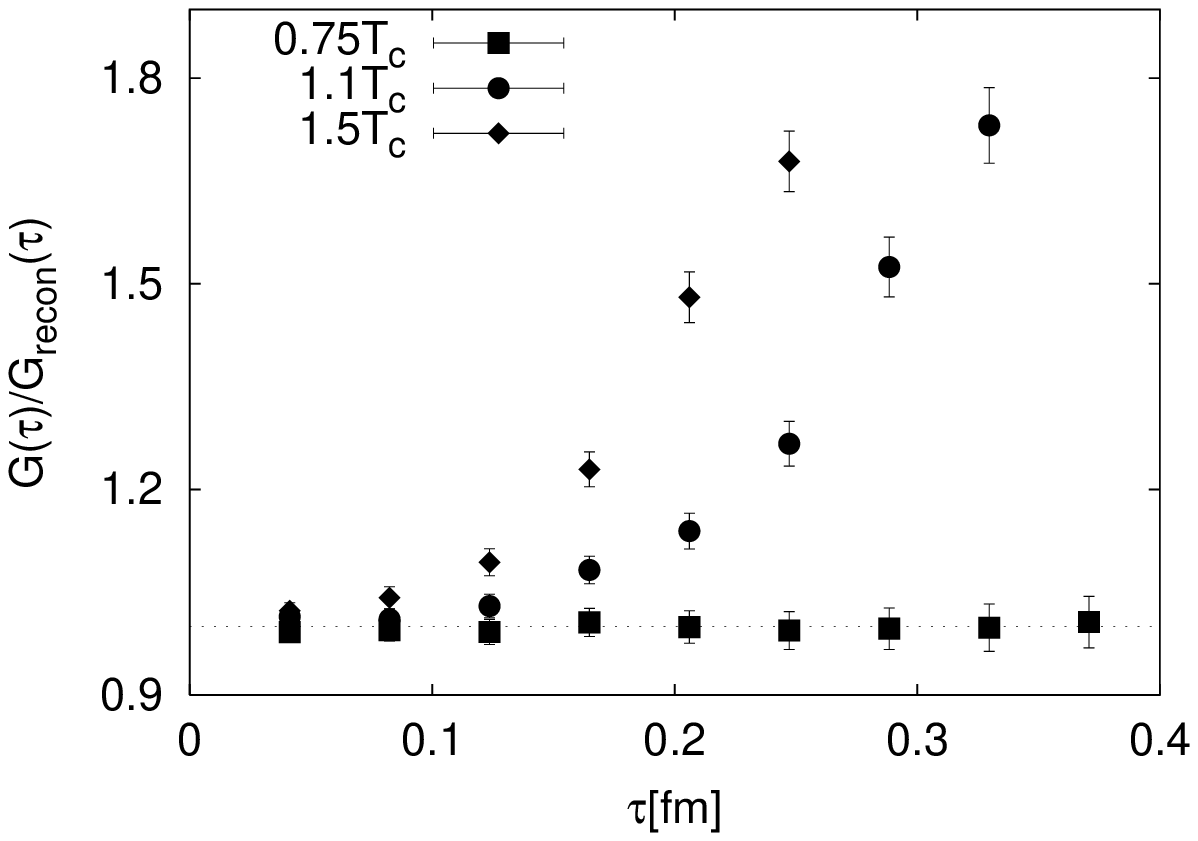,height=53mm}
\end{minipage}
\hspace*{2cm}
\begin{minipage}[htbp]{5.5cm}
\epsfig{file=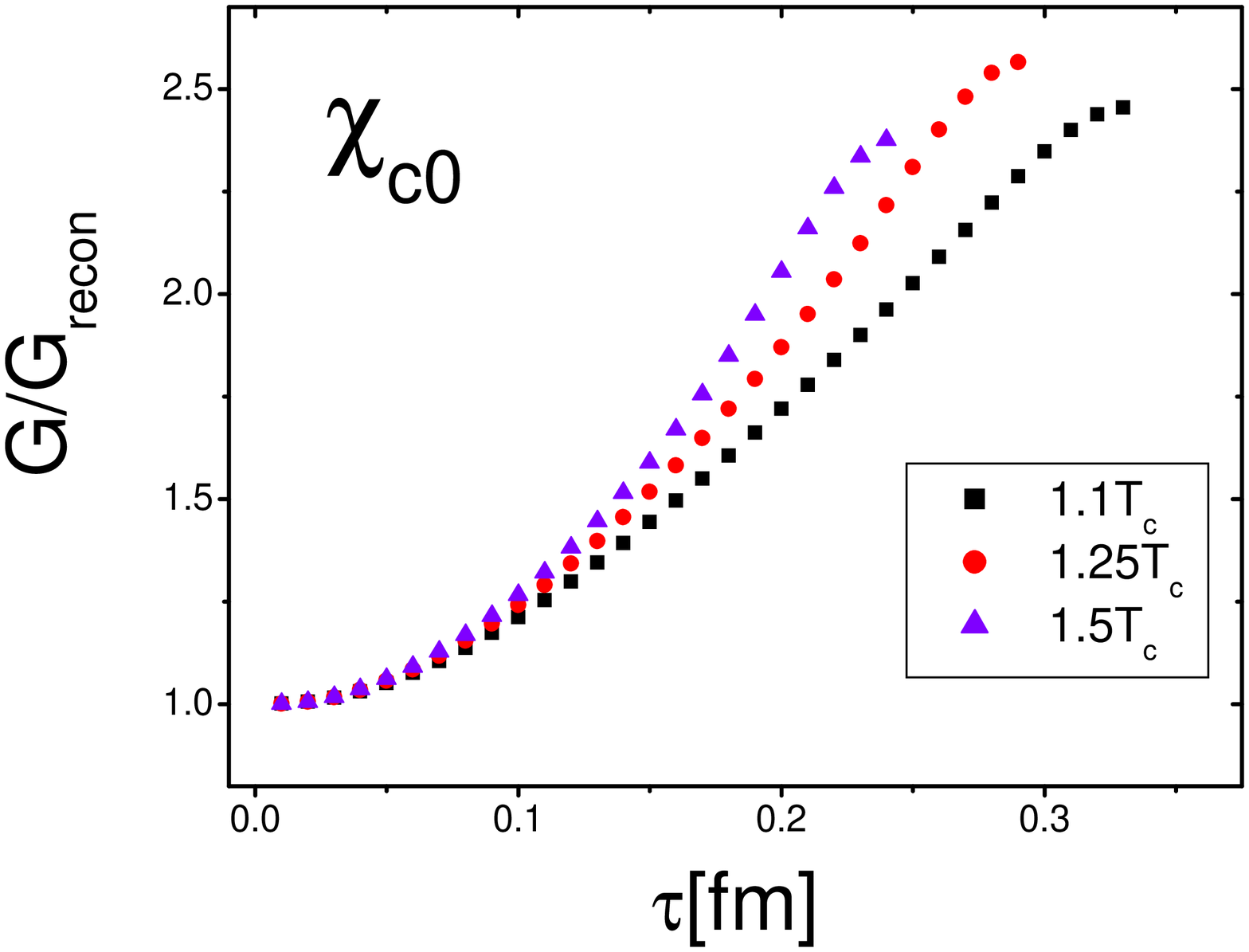,height=53mm}
\end{minipage}
\caption{Ratio of the scalar charmonia correlator to the
reconstructed correlator calculated on the lattice (left panel
from \cite{Datta:2003ww}) and in our model with sharp continuum
threshold (right panel).} \label{fig:sc}
\end{figure}

We present the temperature dependence of the scalar charmonia
correlator, normalized to its reconstructed correlator in
Fig.~\ref{fig:sc}. The left panel displays the results obtained on
the lattice \cite{Datta:2003ww}. These show a very large increase
of the correlator at $1.1T_c$, suggesting that the properties of
the $\chi_{c0}$ are already modified near the transition
temperature. The potential model calculation using (\ref{pot}) as
the potential, and a sharp threshold, i.e.~$f(\omega,s_0)=a_H$ in
(\ref{spft}), with $a_H=-1$ for the scalar, and $1$ for the
pseudoscalar, are presented in the right panel.

The calculated $\chi_{c0}$ correlator shows an increase too, in
qualitative agreement with the lattice data. Despite the fact that
the contribution from the $\chi_{c0}$ state becomes negligible,
the scalar correlator above deconfinement is enhanced compared to
the zero temperature correlator. This enhancement is due to the
thermal shift of the continuum threshold $s_0(T)$.

\begin{figure}[htbp]
\begin{minipage}[htbp]{5.5cm}
\epsfig{file=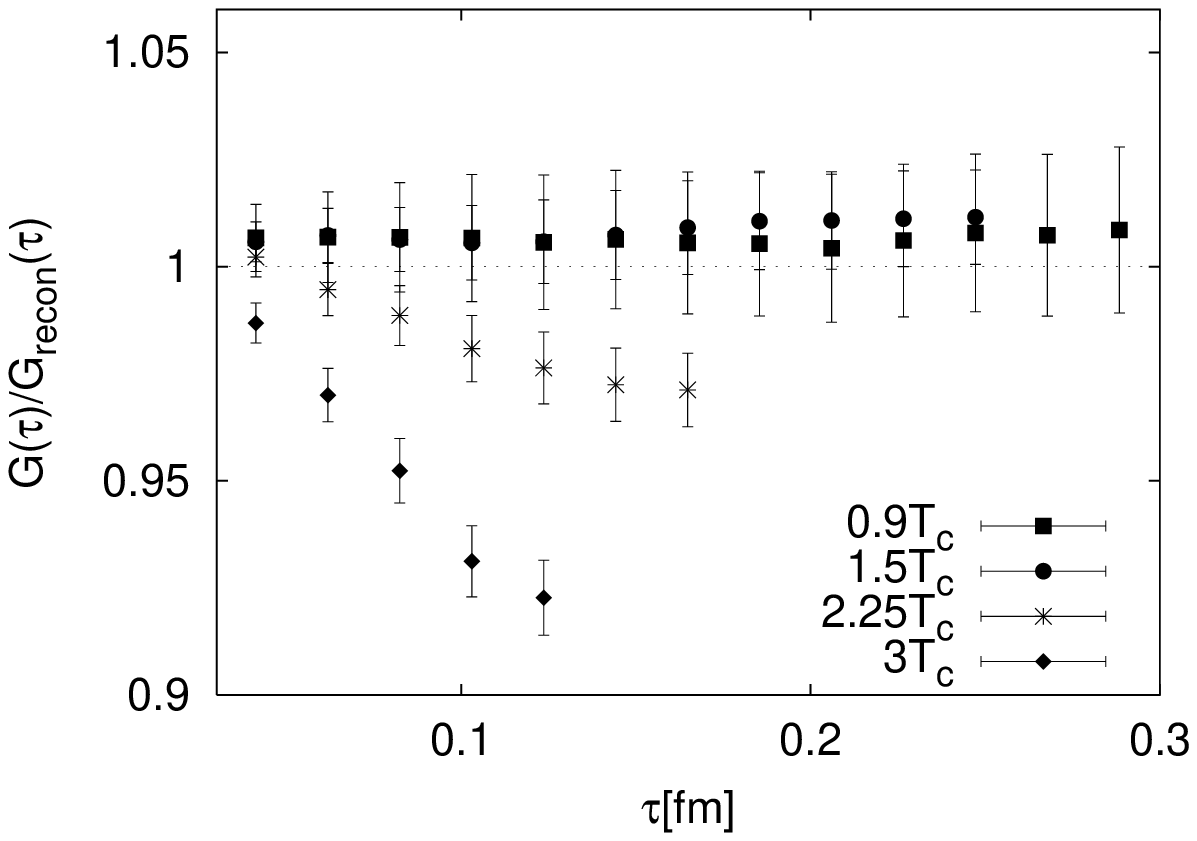,height=53mm}
\end{minipage}
\hspace*{2cm}
\begin{minipage}[htbp]{5.5cm}
\epsfig{file=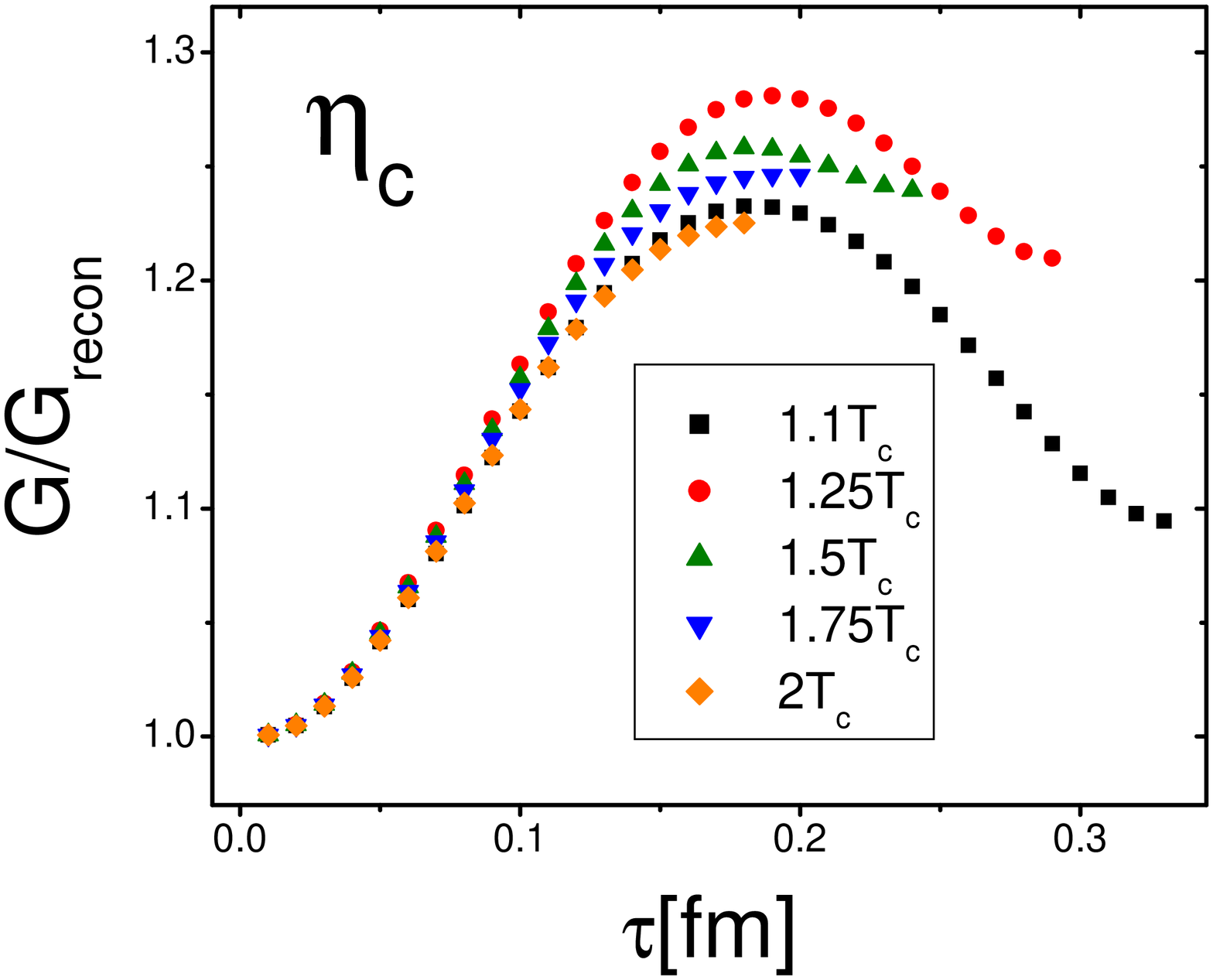,height=53mm}
\end{minipage}
\caption{Ratio of the pseudoscalar charmonia correlator to the
reconstructed correlator calculated on the lattice (left panel
from \cite{Datta:2003ww}) and in our model with sharp continuum
threshold (right panel).} \label{fig:psc}
\end{figure}

The pseudoscalar charmonia correlators are presented in Figure
\ref{fig:psc}. The lattice $\eta_c$ correlator (left panel) shows
no change until about $3T_c$, where a decrease is detected. The
potential model with a sharp threshold (right panel), however,
yields a moderate increase in the correlator at about $0.1~$fm.
This feature is again attributed to the decrease of the continuum
threshold with temperature, and is not detected on the lattice.
After reaching a maximum $G/G_{recon}$ drops, due to the decrease
of the amplitude $F_{PS}(T)$ above deconfinement (c.f.
Fig.~\ref{fig:mass-amp}).

\begin{figure}[htbp]
\begin{minipage}[htbp]{5.5cm}
\epsfig{file=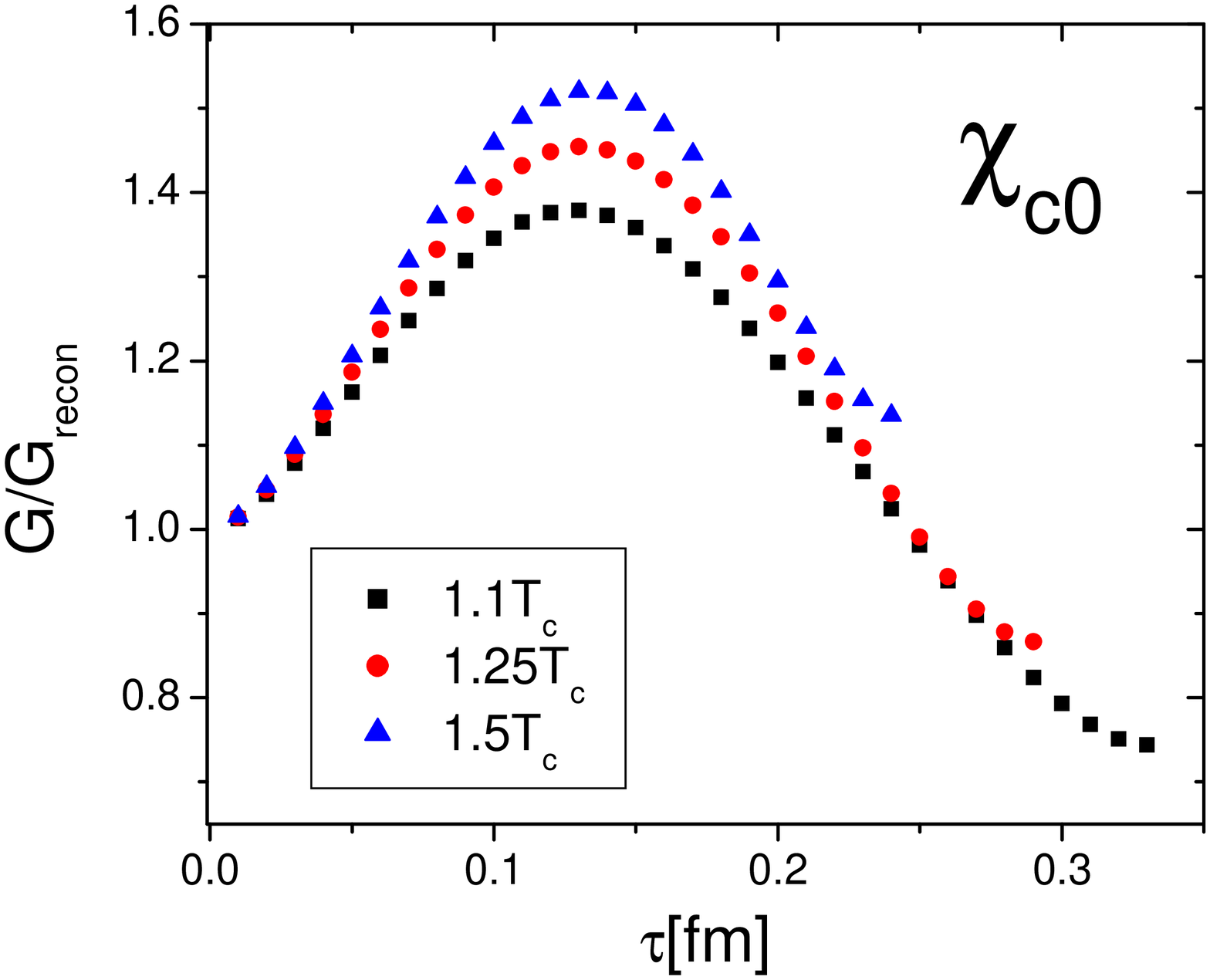,height=53mm}
\end{minipage}
\hspace*{2cm}
\begin{minipage}[htbp]{5.5cm}
\epsfig{file=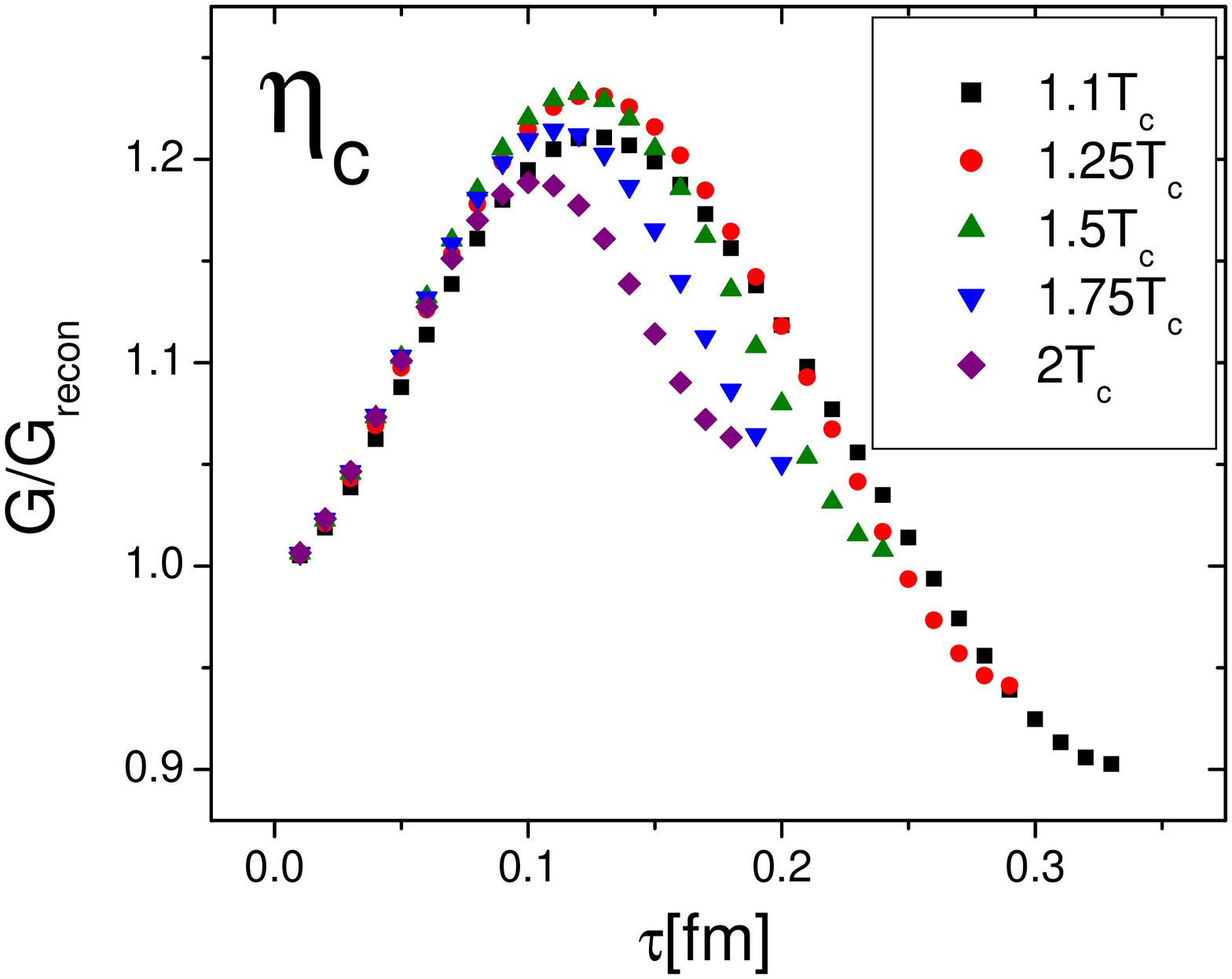,height=53mm}
\end{minipage}
\caption{Ratio of scalar (left panel) and pseudoscalar (right
panel) correlator to reconstructed correlator with smooth
continuum threshold.} \label{fig:smooth}
\end{figure}
We now discuss the results obtained with the smooth continuum
threshold (\ref{smooth}). The temperature dependence of the
$\chi_{c0}$ correlator obtained with the smooth threshold shows a
somewhat different behavior, as seen in the left panel of
Fig.~\ref{fig:smooth}. After enhancement at intermediate $\tau$
the correlator falls below the zero temperature one. We thus see
that the temperature dependence of the scalar correlator is
strongly affected by the continuum part of the spectral function.
In this case too, the reduction of the continuum threshold clearly
leads to the increase of the correlator at intermediate $\tau$.
When comparing the behavior of the $\eta_c$ correlator calculated
with the sharp threshold (right panel in Fig. \ref{fig:psc}) and
with the smooth threshold (\ref{smooth}) (right panel of
Fig.~\ref{fig:smooth}), we conclude that in the pseudoscalar
channel although quantitatively the results are different, for the
qualitative behavior only the reduction of the threshold, and not
the exact form of the continuum matters.

\begin{figure}[htbp]
\begin{minipage}[htbp]{5.5cm}
\epsfig{file=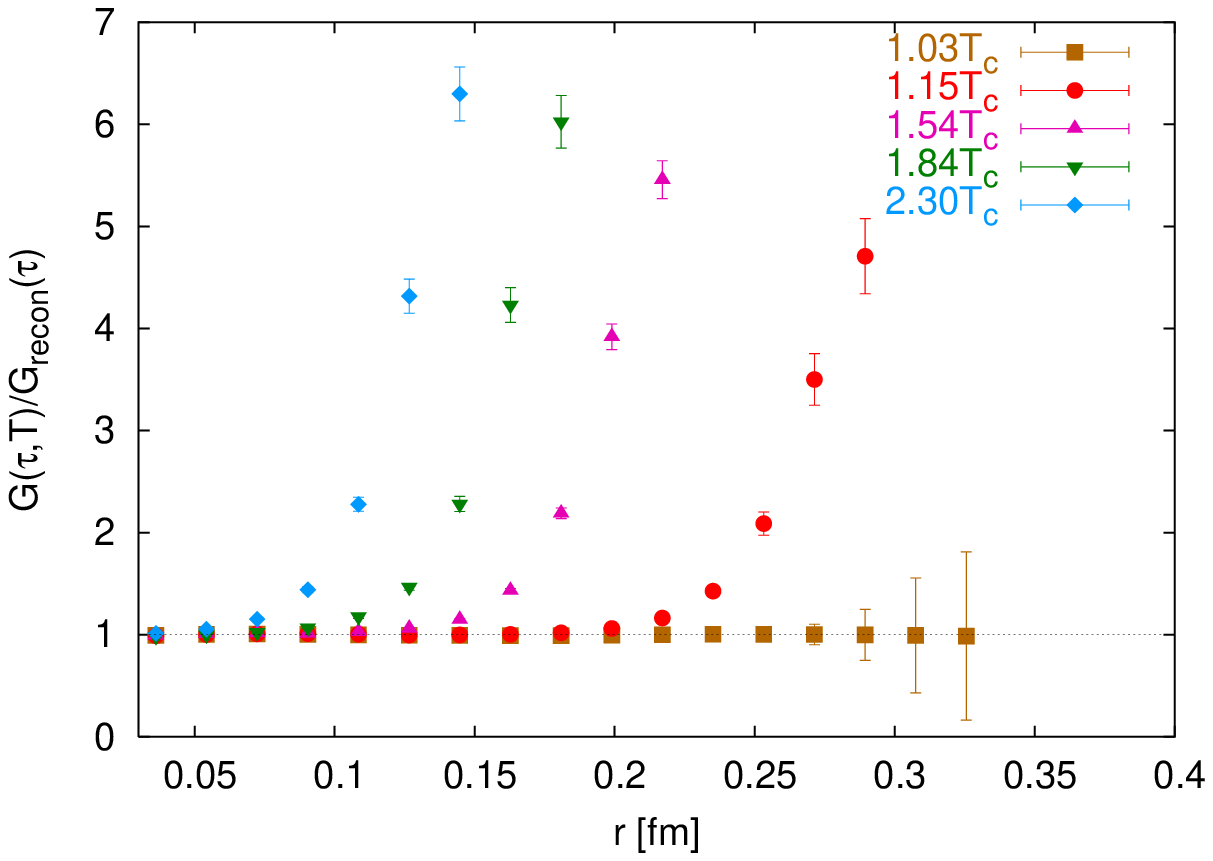,height=53mm}
\end{minipage}
\hspace*{2cm}
\begin{minipage}[htbp]{5.5cm}
\epsfig{file=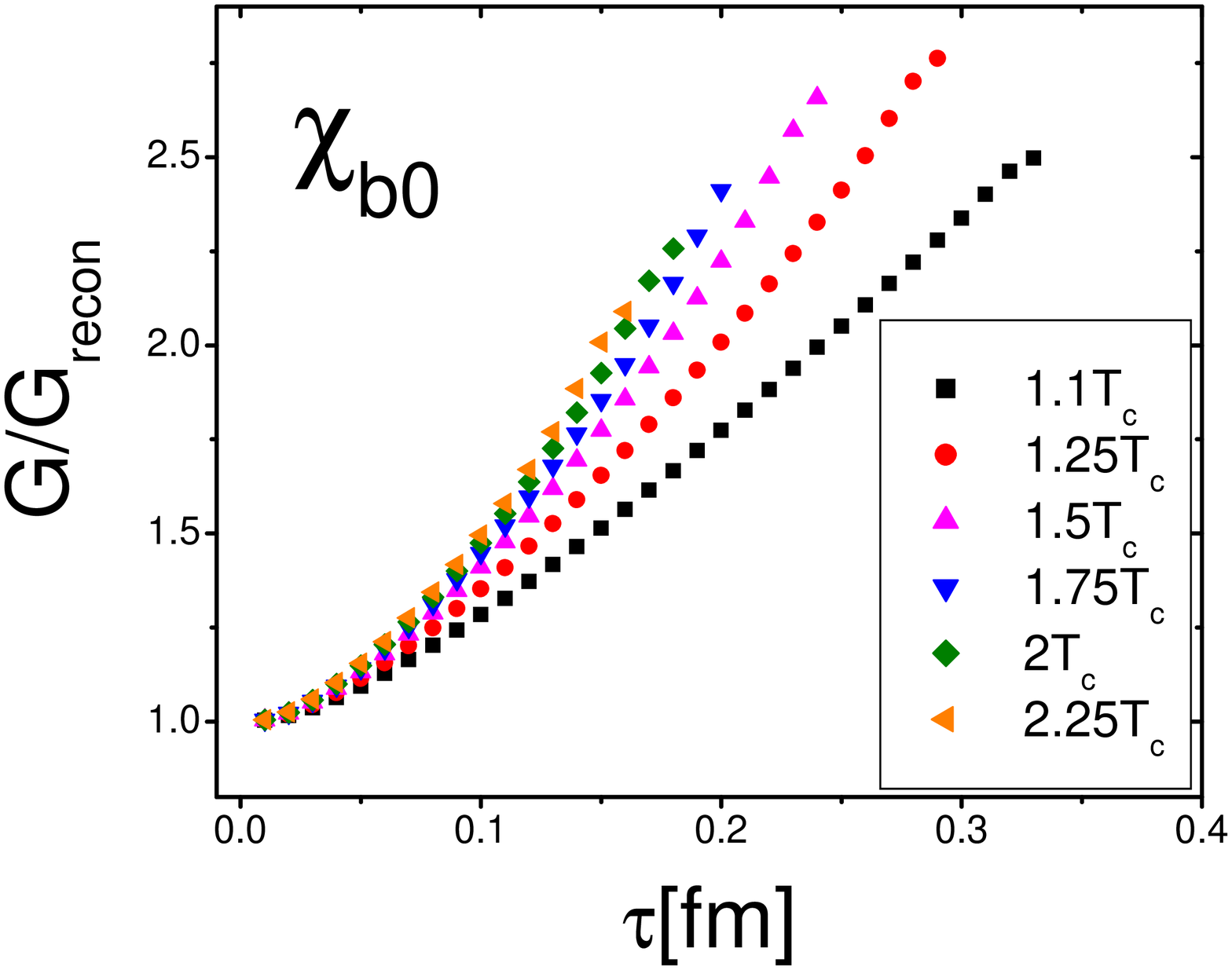,height=53mm}
\end{minipage}
\caption{Ratio of the scalar bottomonia correlator to the
reconstructed correlator calculated on the lattice (left panel
from \cite{Petrov:2005ej}) and in our model with sharp continuum
threshold (right panel).} \label{fig:bs}
\end{figure}
Qualitatively similar behavior was obtained for the bottomonia
states. The scalar bottomonium $\chi_{b0}$ correlator shown in
Fig.~\ref{fig:bs} has an increase already at $1.13T_c$, as
determined both on the lattice (left panel from
\cite{Petrov:2005ej}) and in our model calculations (right panel).
Here the sharp continuum has been used. Thus, the behavior of the
scalar bottomonium channel is very similar to that of the scalar
charmonium, even though, contrary to the $\chi_{c0}$ state, the
$\chi_{b0}$ survives until much higher temperatures. We explain
this with the fact, that the shifted continuum gives the dominant
contribution to the scalar correlator.

\begin{figure}[htbp]
\begin{minipage}[htbp]{5.5cm}
\epsfig{file=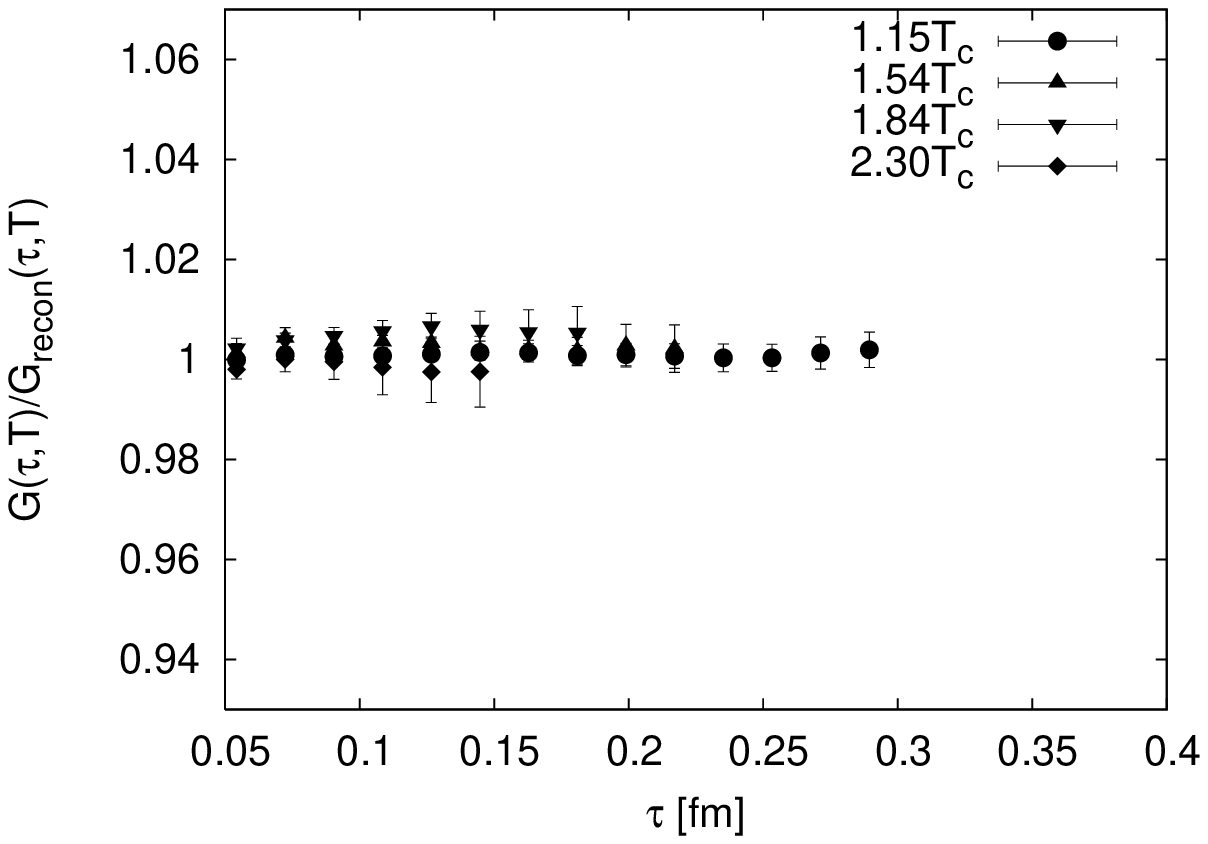,height=53mm}
\end{minipage}
\hspace*{2cm}
\begin{minipage}[htbp]{5.5cm}
\epsfig{file=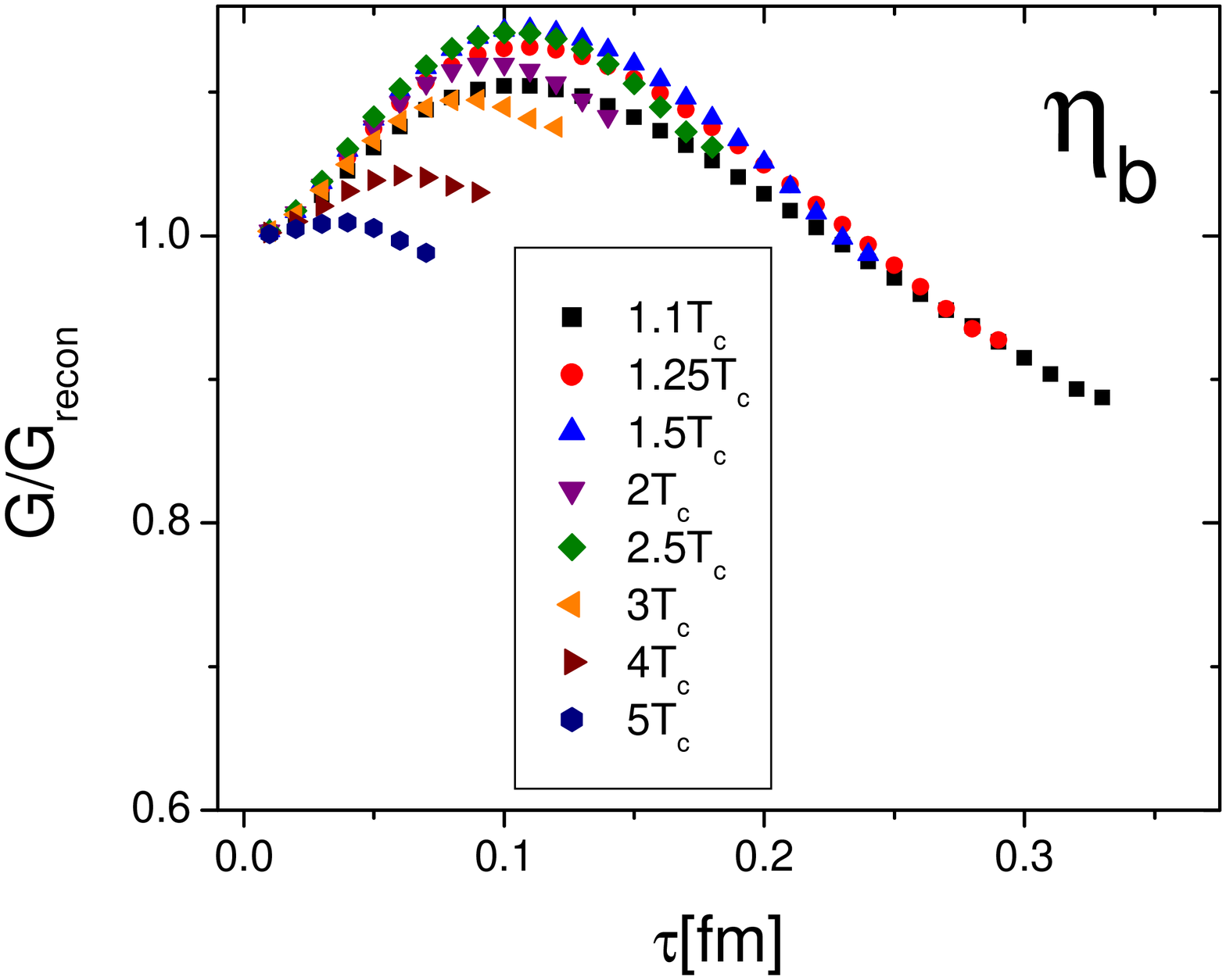,height=53mm}
\end{minipage}
\caption{Ratio of the pseudoscalar bottomonia correlator to the
reconstructed correlator calculated on the lattice (left panel
from \cite{Petrov:2005ej}) and in our model with sharp continuum
threshold (right panel).} \label{fig:bps}
\end{figure}

In Fig.~\ref{fig:bps} the behavior of the pseudoscalar bottomonium
correlator is illustrated. Lattice results for this channel (left
panel) show no deviation from one in the correlator ratio
$G/G_{recon}$ up to high temperatures. This is considered to be an
indication of the temperature independence of the $\eta_b$
properties up to these temperatures. The potential model studies
for the pseudoscalar correlator (right panel), yield again, a
qualitatively different behavior than seen from the lattice. There
is an increase at small $\tau$, and a drop at large $\tau$ in the
correlator compared to the zero temperature correlator. The
increase is due to the threshold reduction, and the decrease is
due to the reduction of the amplitude.

In summary, in this Section we have seen that $G/G_{recon}$ for
the pseudoscalar channel is increasing above one as the result of
the shift in the continuum threshold, and its decrease is due to
the decrease in the amplitude $F_{PS}$. This result is independent
of the detailed form of the continuum. The temperature dependence
of the scalar correlator is, on the other hand, sensitive to the
continuum part of the spectral function.

%%%%%%%%%%%%%%%%%%%%%%%%%%%%%%%%%%%%%%%%%%%%%%%
\subsection{Results with Higher Excited States}
\label{excited}

\begin{figure}[htbp]
\begin{minipage}[htbp]{5.5cm}
\epsfig{file=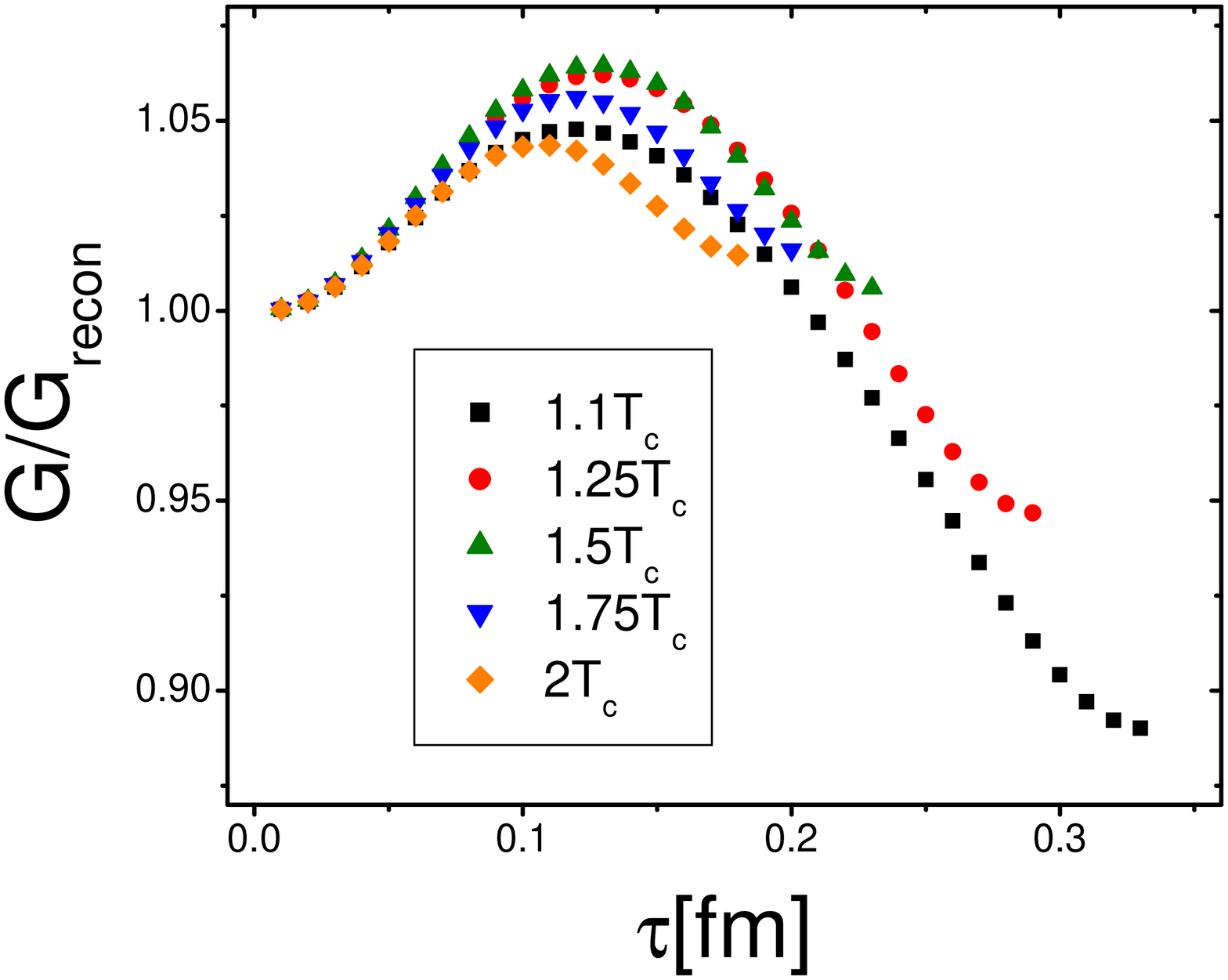,height=53mm}
\end{minipage}
\hspace*{2cm}
\begin{minipage}[htbp]{5.5cm}
\epsfig{file=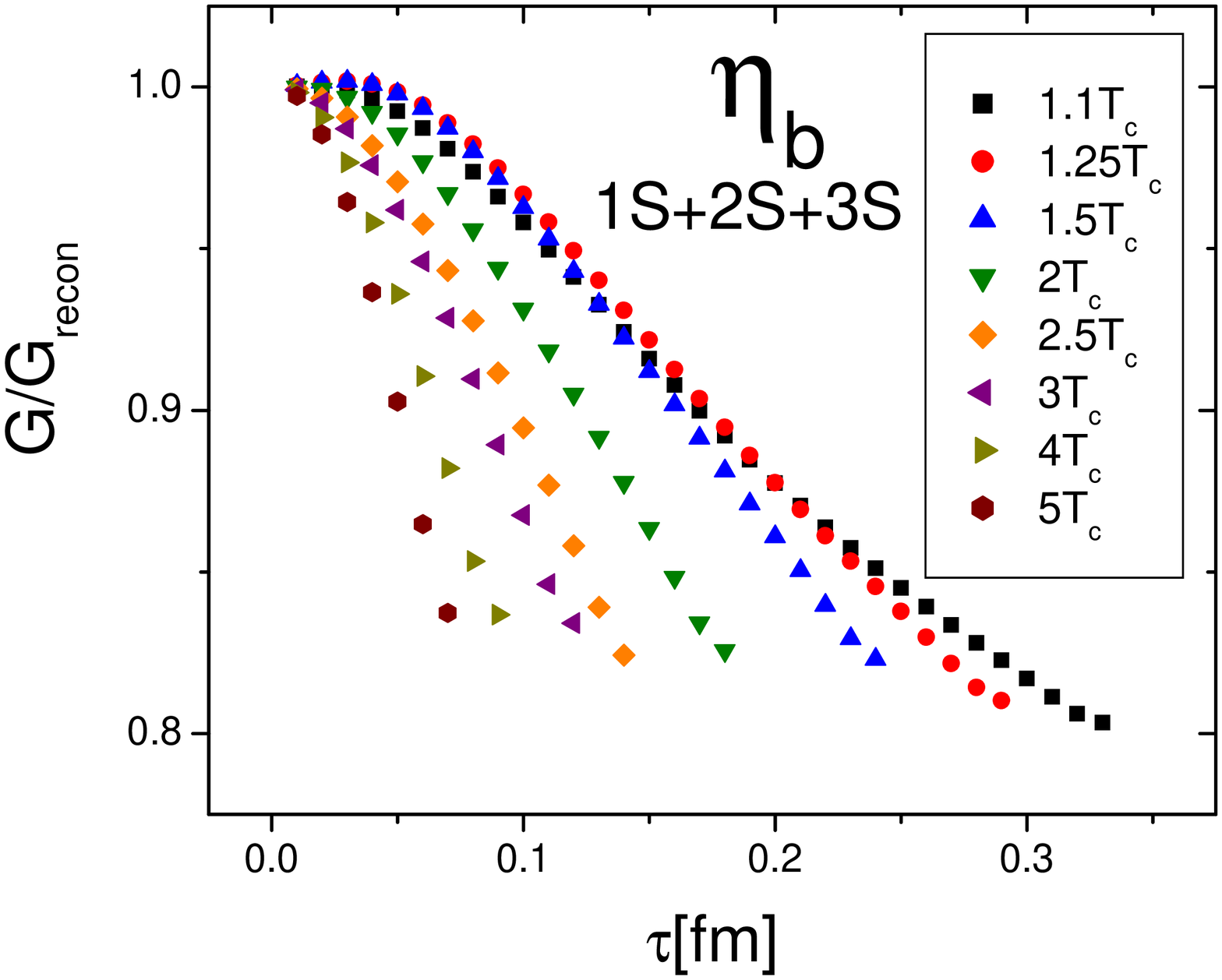,height=53mm}
\end{minipage}
\caption{Ratio of the pseudoscalar correlator to the reconstructed
correlator when including 1S and 2S charmonium states (left panel)
and 1S, 2S and 3S bottomonium states (right panel).}
\label{fig:seta}
\end{figure}

In the analysis presented so far we have considered only the
lowest meson states in a given channel. We have done so, because
we wanted to make contact with the existing lattice data, in which
the excited states in any given channel were not yet identified.
In order to study possible effects due to the higher states in a
given channel, we now include also the 2S state, and the 2S and 3S
states for the pseudoscalar charmonium, and bottomonium,
respectively. These states enter directly in the first term of the
spectral function (\ref{spft}). Since the 2S charmonium state is
melted at $T_c$ (see Table \ref{tab:c} in Appendix
\ref{app:tables}), we expect a larger drop in $G/G_{recon}$
compared to the previous case, where only the 1S state is
included. Similarly, the 3S bottomonium state is melted at $T_c$,
but the 2S state survives up until about $1.75T_c$ (see Table
\ref{tab:b} in Appendix \ref{app:tables}). The results are
presented in Fig.~\ref{fig:seta}. The left panel illustrates the
temperature dependence of the $\eta_c$ correlator obtained when
both the 1S and the 2S states are accounted for. As expected, we
identify a 10\%-20\% reduction in the correlator, attributed to
the melting of the 2S state. The lattice data on the pseudoscalar
correlator shows no evidence for such a drop. Instead,
$G/G_{recon}\simeq 1$ within errors, up to $2.25T_c$. The right
panel in Fig.~\ref{fig:seta} shows the $\eta_b$ correlator, where
the melting of the 3S state near the transition, and the 2S state
at higher temperatures, clearly induces a 10\%-20\% drop of the
correlator compared to the case with only the 1S state included.

%%%%%%%%%%%%%%%%%%%%%%%%%%%%%%%%%%%%%%%%%%%%%%
\subsection{Results Using the Internal Energy}

\begin{figure}[h]
\begin{minipage}[h]{5.5cm}
\epsfig{file=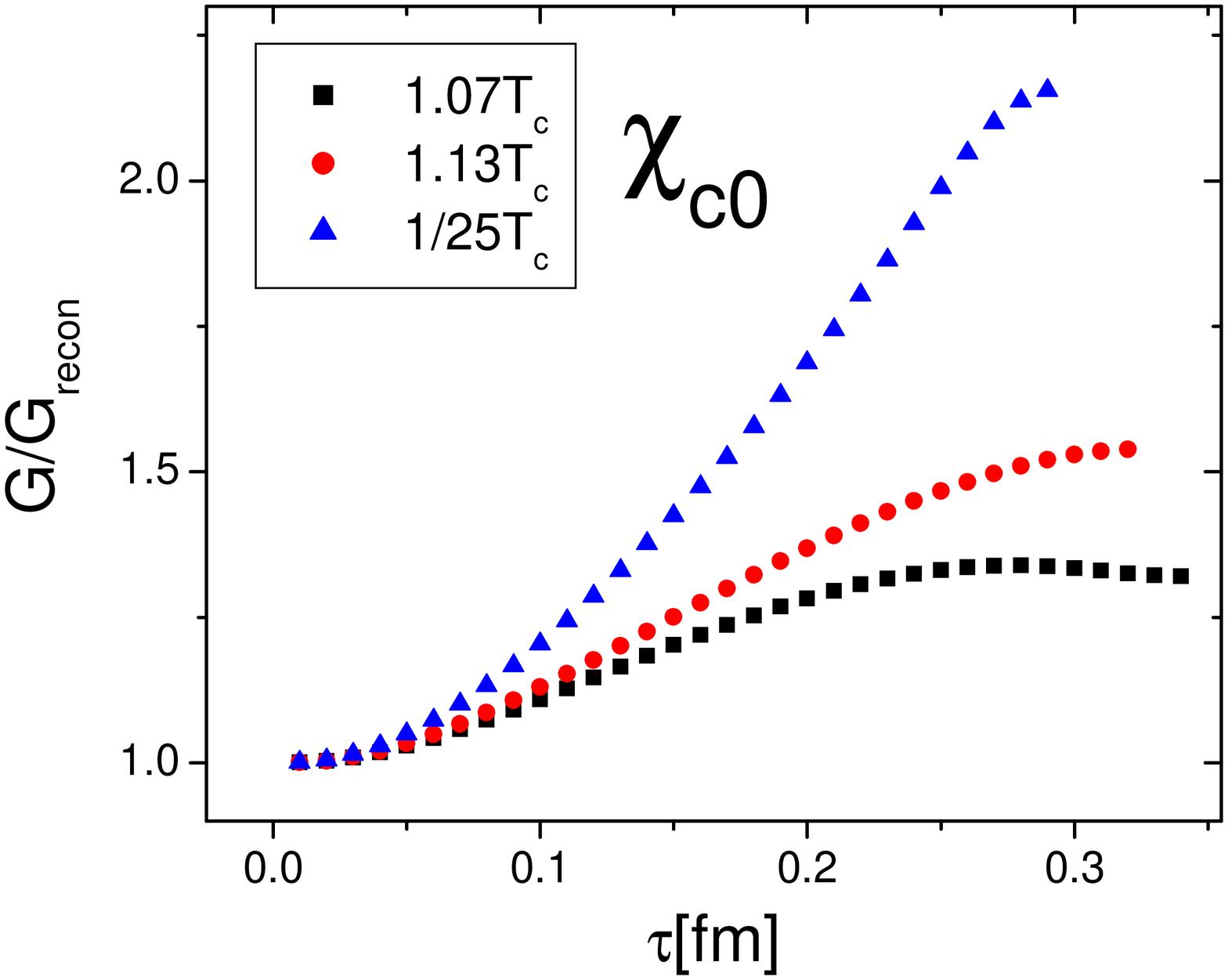,height=53mm}
\end{minipage}
\hspace*{2cm}
\begin{minipage}[h]{5.5cm}
\epsfig{file=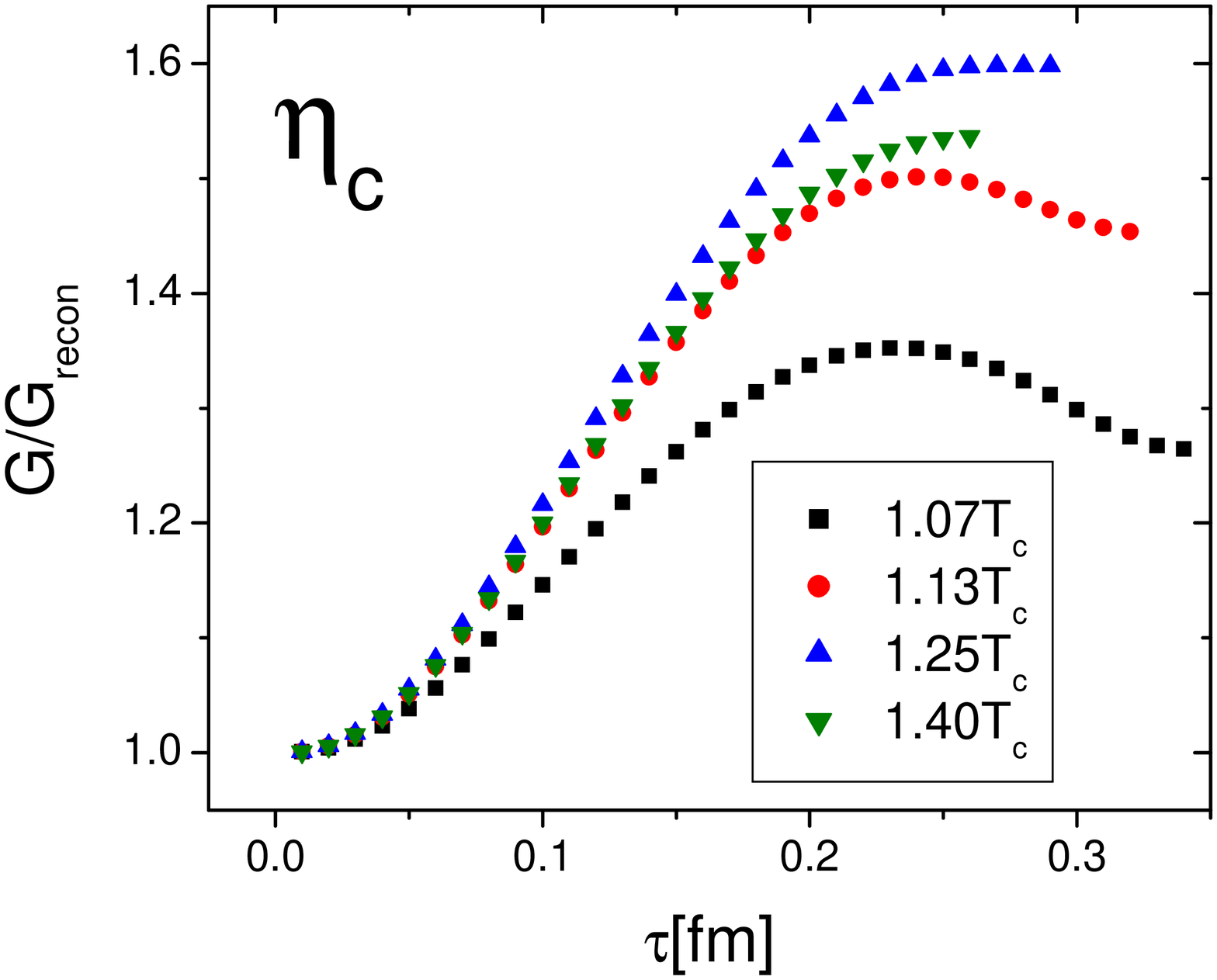,height=53mm}
\end{minipage}\\
\begin{minipage}[h]{5.5cm}
\epsfig{file=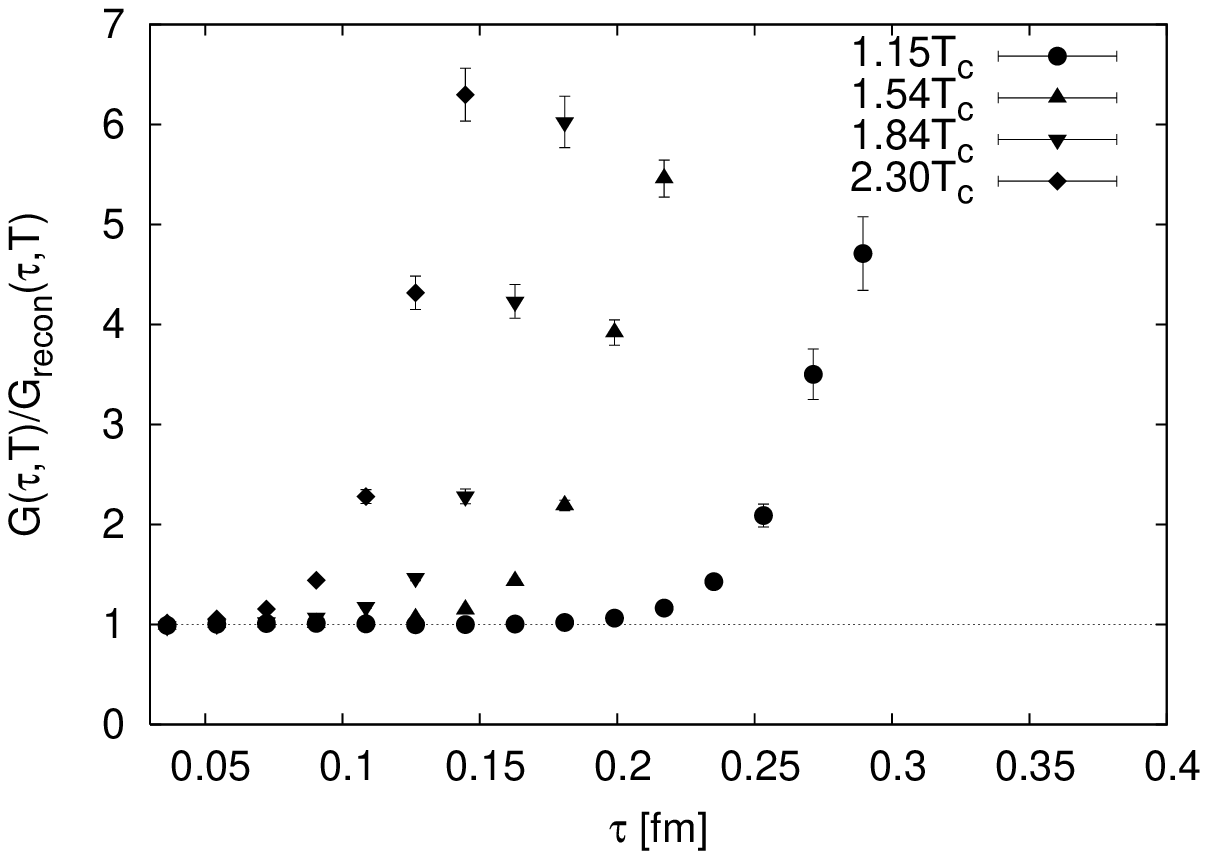,height=53mm}
\end{minipage}
\hspace*{2cm}
\begin{minipage}[h]{5.5cm}
\epsfig{file=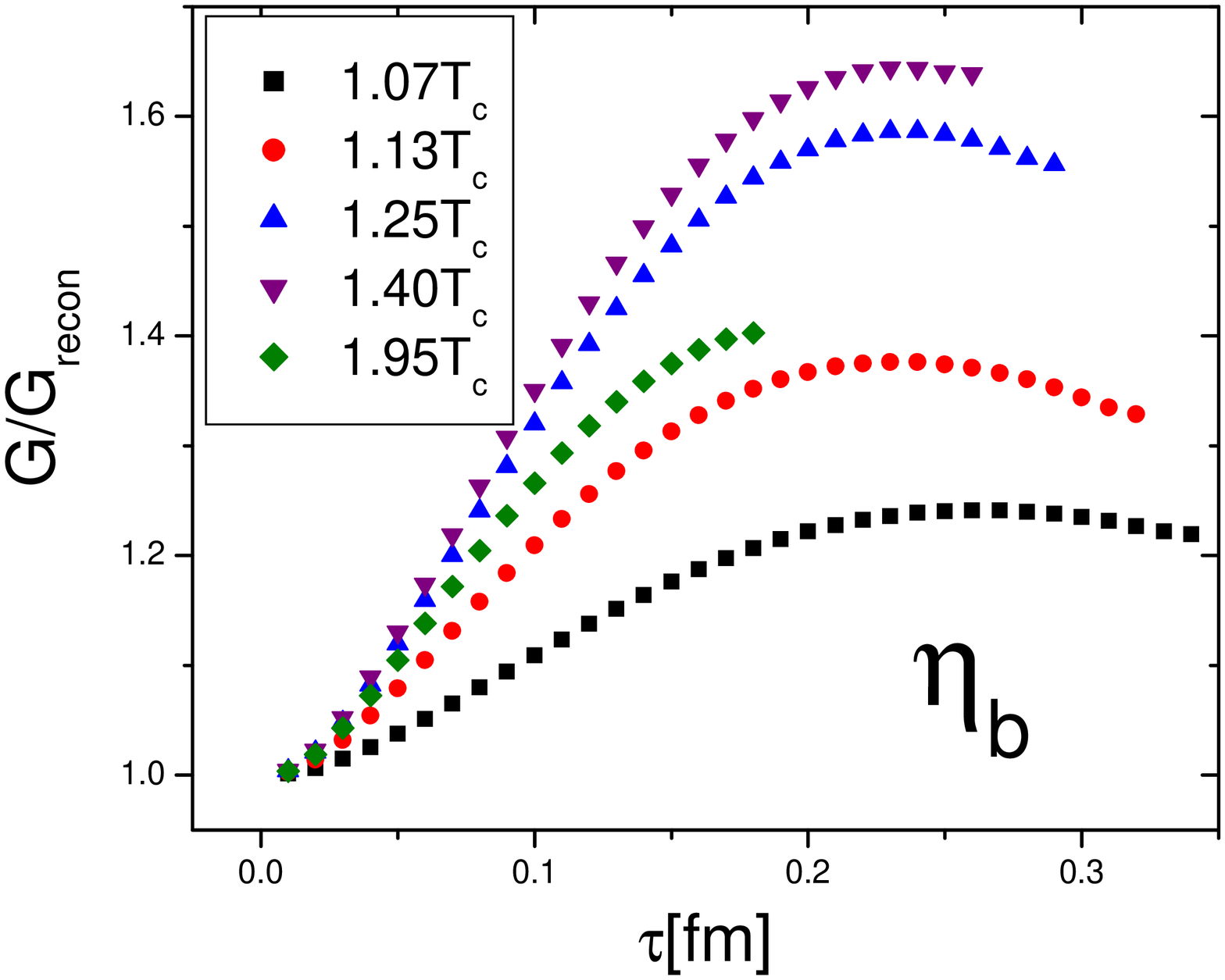,height=53mm}
\end{minipage}
\caption{Ratio of charmonia (top panels) and bottomonia (bottom
panels) correlators to the reconstructed correlators for scalar
(left panels) and pseudoscalar (right panels) channels obtained
with the potential fitted to lattice internal energy and sharp
continuum threshold.} \label{fig:latfit}
\end{figure}

We end this Section with the discussion of the quarkonia
correlators obtained using as the potential the internal energy of
a static quark-antiquark pair determined from the lattice. The
temperature dependence of the correlators is illustrated in
Fig.~\ref{fig:latfit}. In the numerical analysis we used the sharp
continuum. As one can see from the left panels of
Fig.~\ref{fig:latfit} the behavior of the scalar correlator is
similar to the one obtained in Section \ref{cornell}: Even though
the $\chi_{c0}$ and the $\chi_{b0}$ states are dissociated, above
$T_c$ the correlator is always enhanced relative to the zero
temperature one. As shown on the right panels of
Fig.~\ref{fig:latfit} the enhancement in the pseudoscalar channel
is larger than for the screened Cornell potential discussed in
Section \ref{cornell}. As mentioned in Section \ref{res-prop}, the
masses and amplitudes of the 1S quarkonia states calculated using
the internal energy as the potential, show a significant increase
at $T>T_c$. The increase in the amplitudes translate into a
significant enhancement of the correlators with respect to their
values determined with the zero temperature amplitudes. Besides
this increase due to the amplitude, our results prove to be
qualitatively insensitive to the detailed form of the potential,
suggesting that they are based on very general physical arguments.
Thus we tested the robustness of our results regarding the
discrepancy between potential model studies and lattice results.
%

%%%%%%%%%%%%%%%%%%%%%%%%%%%%%%%%%%%%%%%%%%%%%%%%%%%%%
\section{Numerical Results for the Vector correlator}
\label{res-corrII}

\begin{figure}[htbp]
\begin{minipage}[htbp]{5.5cm}
\epsfig{file=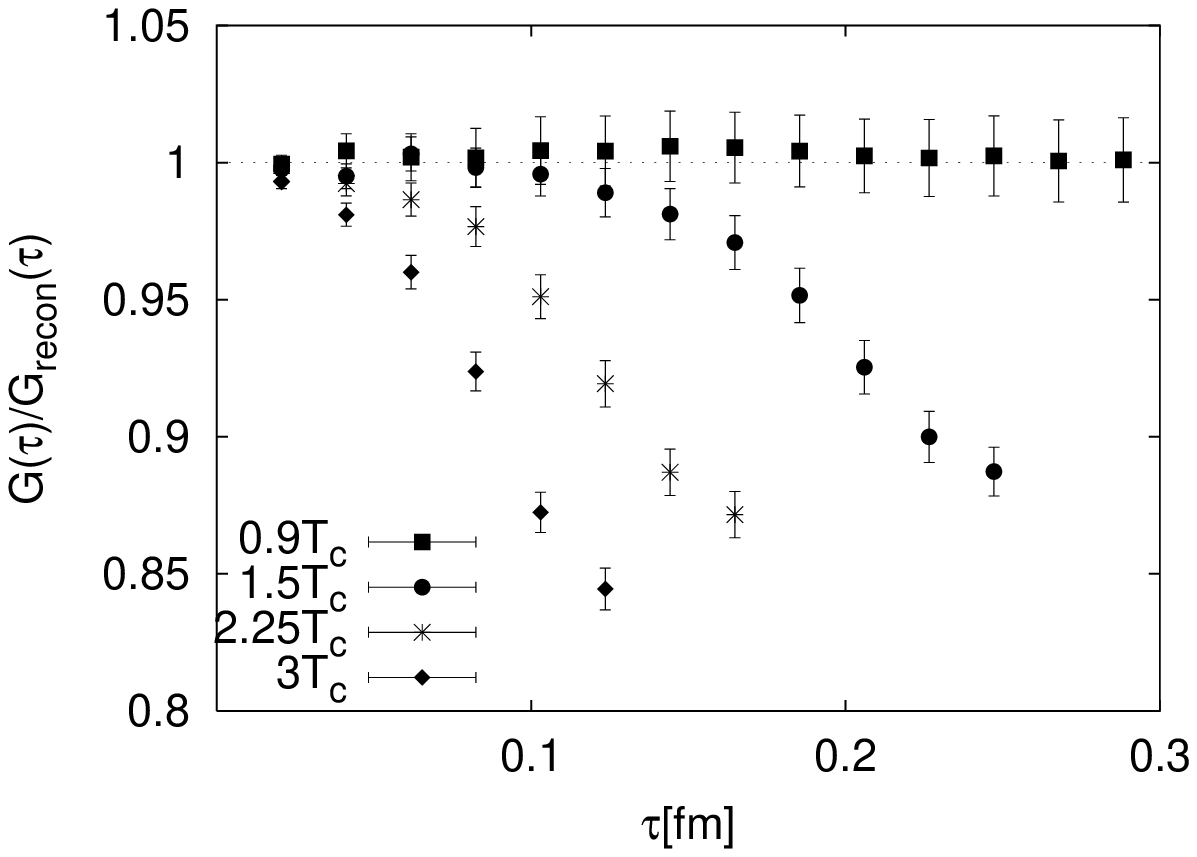,height=53mm}
\end{minipage}
\hspace*{2cm}
\begin{minipage}[htbp]{5.5cm}
\epsfig{file=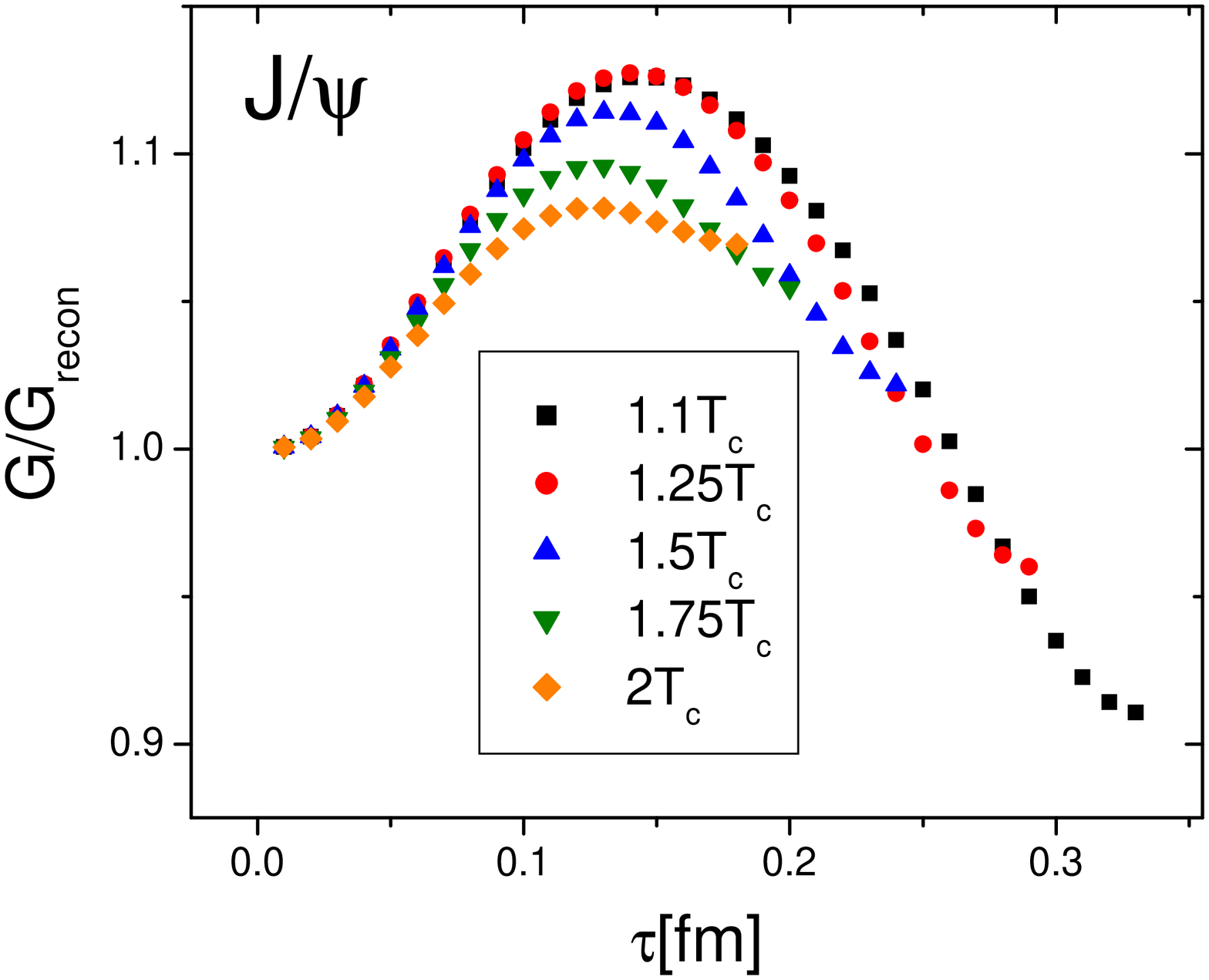,height=53mm}
\end{minipage}
\caption{Ratio of the $J/\psi$ correlator to the reconstructed
correlator calculated on the lattice (left panel from
\cite{Datta:2003ww}) and in our model with sharp continuum
threshold (right panel). } \label{fig:jpsi}
\end{figure}

The vector correlator corresponds to an especially interesting
channel. This is because of the fact, that the vector current is
conserved. As mentioned in Section \ref{model} (for the detailed
analysis see Appendix \ref{transport}), this leads to an extra
contribution in the spectral function compared to the scalar and
pseudoscalar channels. This contribution (the third term in
equation (\ref{spft})) arises from the diffusion and charge
fluctuations specific only for the vector state. Also, lattice
data shows a difference in the temperature dependence of the
pseudoscalar and the vector correlators. Namely, in the
pseudoscalar channel the ratio $G/G_{recon}$ is equal to one up to
temperatures $\sim 2.25 T_c$, and shows significant deviation at
about $3T_c$, as illustrated on the left panel of
Fig.~\ref{fig:psc}. In the vector channel, shown in the left panel
of Fig.~\ref{fig:jpsi}, this ratio significantly decreases from
one already at $1.5T_c$ at distances $>0.15~$fm. This figure
further illustrates, that with increasing temperature the
deviation happens already at smaller distances. Since to leading
order in the non-relativistic expansion the pseudoscalar and
vector channels correspond to the same 1S state, one would not
expect the correlator of the $\eta_c$ and the $J/\psi$ to behave
differently. Here we conjecture, that the effects of diffusion and
charge fluctuations make the $J/\psi$ correlator smaller than the
correlator of the $\eta_c$.

The right panel of Fig.~\ref{fig:jpsi} displays the $J/\psi$
correlator from our potential model calculations. We see a rise of
the correlator at small distances, where the continuum
contribution to the spectral function is dominant, and thus again,
the reduction of the threshold is manifest. As in the case of the
pseudoscalar, the model calculations do not reproduce the behavior
of the vector correlator obtained from the lattice. When comparing
the $\eta_c$ and the $J/\psi$ correlator from the model
calculations, i.e.~the right panels of Figs.~\ref{fig:psc} and
\ref{fig:jpsi}, at large distances we identify a reduction in the
vector channel compared to the pseudoscalar channel, that results
from diffusion and charge fluctuations.
\begin{figure}[htbp]
\begin{minipage}[htbp]{5.5cm}
\epsfig{file=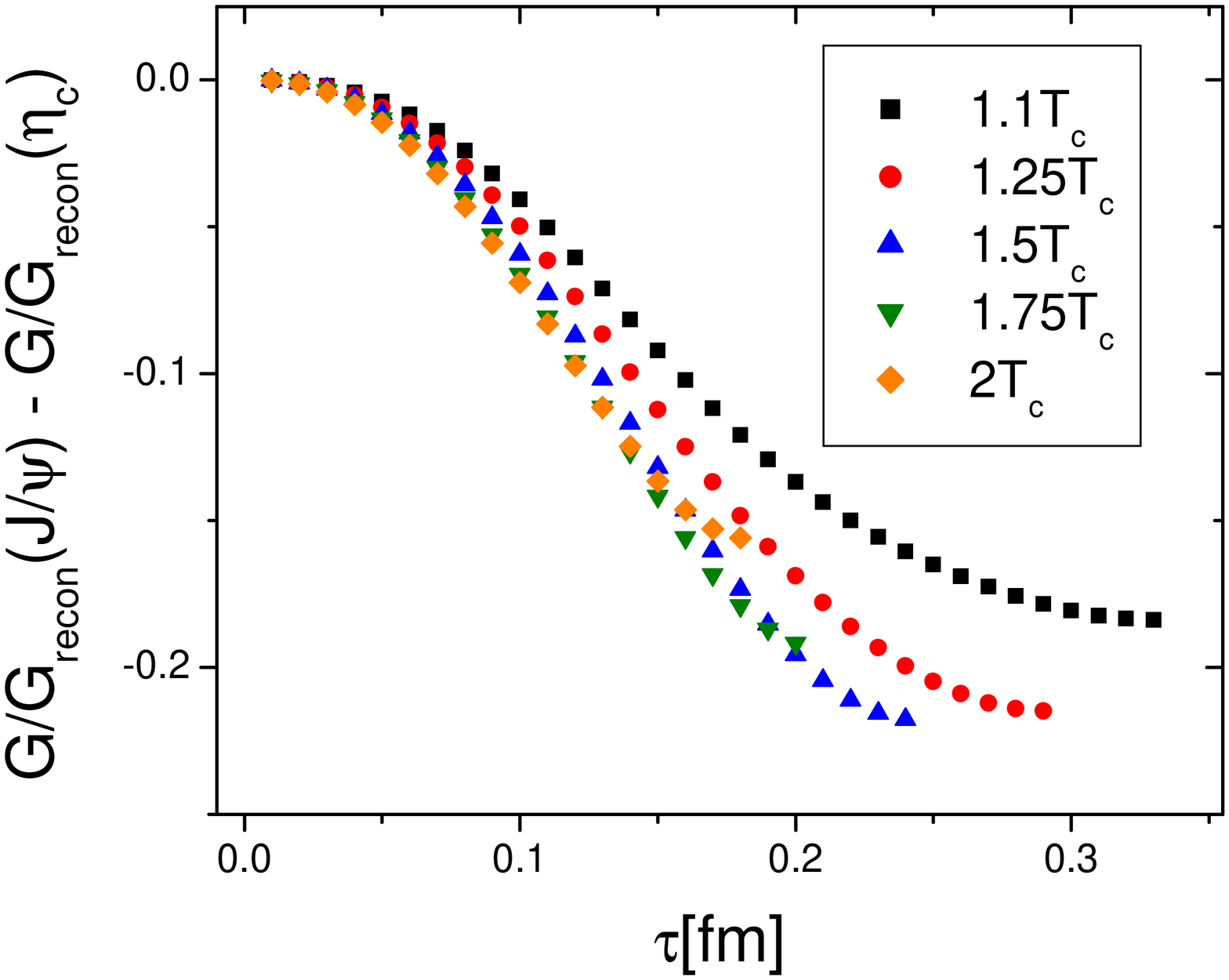,height=53mm}
\end{minipage}
\hspace*{2cm}
\begin{minipage}[htbp]{5.5cm}
\epsfig{file=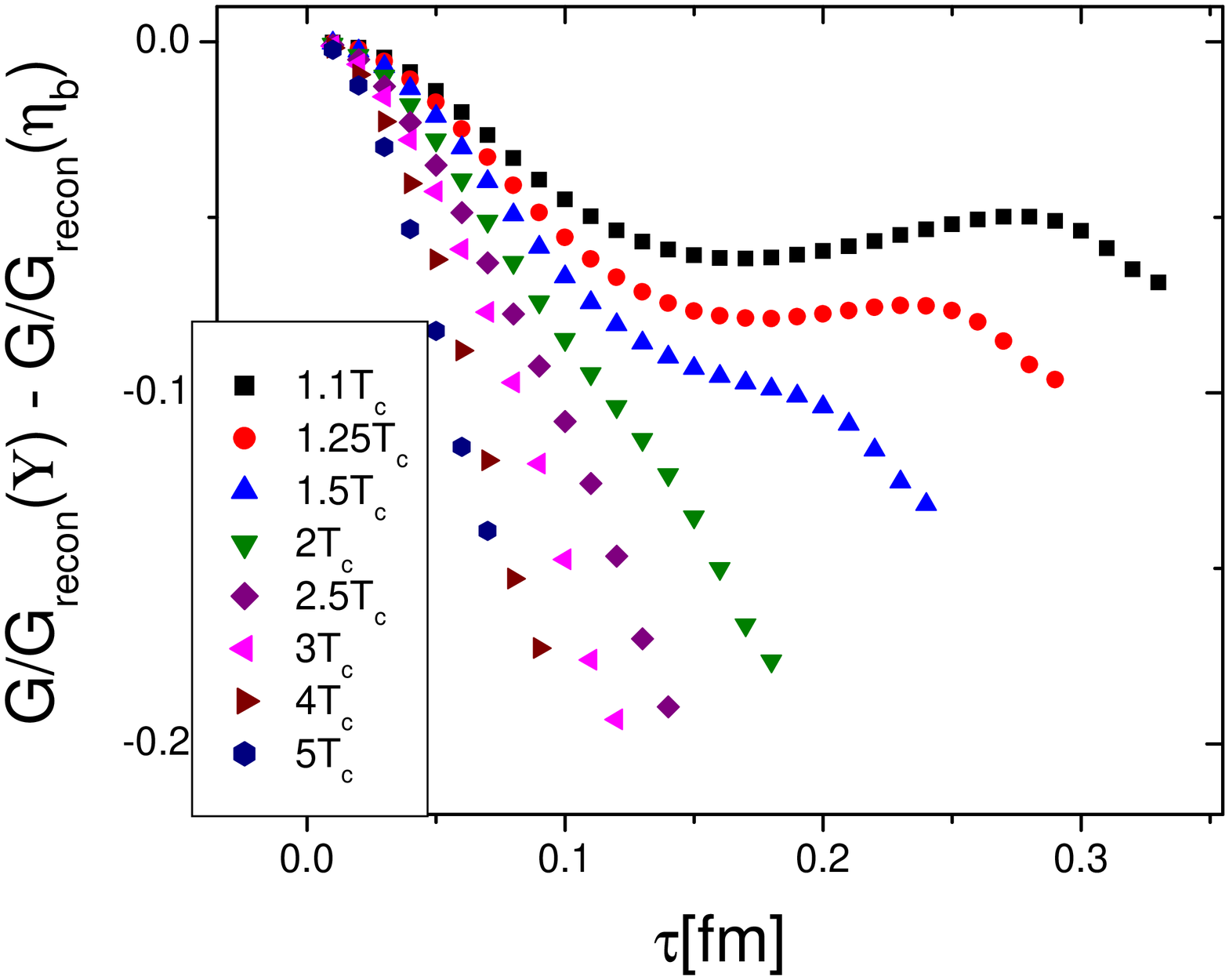,height=53mm}
\end{minipage}
\caption{Difference in the ratios of the correlator to
reconstructed correlator in the vector and pseudoscalar channels
for charmonium (left panel) and bottomonium (right panel). }
\label{fig:diff}
\end{figure}
To better visualize the difference between the temperature
dependence of the pseudoscalar and vector correlators, in Figure
\ref{fig:diff} we show the difference between the corresponding
$G/G_{recon}$ ratios for the charmonium (left panel) and the
bottomonium (right panel) cases. From this figure one can clearly
identify the $20\%$ drop in the $J/\psi$ correlator relative to
the $\eta_c$ correlator. Similar to the charmonium case, for the
bottomonium we also see a reduction at long distances of the
$\Upsilon$ correlator compared to the $\eta_b$ one. As expected,
this reduction is more pronounced with increasing temperature, and
for a given temperature is smaller for bottomonium than for
charmonium.

%%%%%%%%%%%%%%%%%%%%%
\section{Conclusions}
\label{conclude}

In this work, we presented the first detailed study of the
quarkonia properties and correlators using model spectral
functions based on the potential picture with screening, and we
contrasted these with lattice data. While the potential model with
certain screened potentials can reproduce the qualitative features
of the lattice data on the quarkonia spectral functions, namely
the survival of the 1S state up to $1.6T_c$ and dissociation of
the 1P charmonia state at $T \simeq 1.1T_c$, the temperature
dependence of the corresponding meson correlators is different
from that seen on the lattice. In general, the
temperature-dependencies of the correlators show much richer
structure than those calculated on the lattice.

We identified several causes for the difference in the temperature
dependence of the meson correlators: First, the properties of the
different charmonia and bottomonia bound states, such as the
masses and the wave functions, are significantly modified in the
presence of screening, and will change the bound state
contribution to the spectral function. Second, the behavior of the
correlators is affected by the reduction of the continuum
threshold. In a potential model with a screened potential the
continuum threshold is related to the asymptotic value of the
potential, which decreases with increasing temperature. Finally,
the higher excited states (like the 2S, 3S, etc) are expected to
melt at smaller temperatures above $T_c$, inducing a large
temperature dependence for the correlators.

What are the possible implications of these findings for the
picture of quarkonium binding at finite temperature? It is
possible that the effects of the medium on quarkonia binding
cannot be understood in a simple potential model. The effects of
screening, which according to lattice calculations are present in
the plasma \cite{Petreczky:2005bd}, could have time scales which
are not small compared to the typical time scale of the heavy
quark motion. In this case screening cannot affect significantly
the properties of the quarkonia.

On the other hand, the implications of the finite temperature
lattice data on quarkonia correlators also need to be reexamined.
It is known that the high energy part of the quarkonia spectral
functions is strongly influenced by lattice artifacts
\cite{Datta:2003ww}, resulting in the distortion of the continuum
part of the quarkonia spectral functions. Furthermore, the excited
states were not identified in the spectral functions determined on
the lattice. Based on general physical arguments, though, one
would expect the melting of the 2S charmonia and the 3S bottomonia
states close to $T_c$, together with the shift of the continuum,
to induce a stronger temperature dependence of the correlators
than actually observed in the lattice calculations. This statement
is independent of whether or not the medium effects on quarkonia
properties can be properly described in terms of potential models.
In the future, therefore, this problem must be investigated in
more detail.

%%%%%%%%%%%%%%%%%%%%%%%%%%%
\section*{Acknowledgements}

We thank A.~Dumitru, G.~Moore, D.~Teaney for discussions. We thank
S.~Datta for some of the lattice figures, and P.~Sorensen for the
careful reading of the manuscript. This work was partly supported
by U.S. Department of Energy under contract DE-AC02-98CH10886.
During the course of this work A.~M. has been an Alexander von
Humboldt Fellow, and P.~P. has been a Goldhaber and is a RIKEN-BNL
Fellow.

%%%%%%%%%
\appendix
%%%%%%%%%

%%%%%%%%%%%%%%%%%%%%%%%%%%%%%%%%%%%%%%%%%%%%%%%%%%%%%%%%%%%%%%%%%%
\section{Integral representation of the imaginary time correlator}
\label{app-spectral}

Here we derive the spectral representation for the Euclidean
correlator. We start from the Fourier decomposition of the real
time two-point function
\be
D^>(t) = \int_{-\infty}^{\infty} \frac{d\omega}{2\pi} D^>(\omega)
e^{-i\omega t} \, . \ee The Fourier transform is given in terms of
the spectral function $\sigma(\omega)$,
\be
D^>(\omega) = 2\pi\left[1+n(\omega)\right]\sigma(\omega)\,
,\label{Dom}\ee with $n(\omega) = 1/(e^{\beta\omega}-1)$ and
$\beta=1/T~$ \cite{lebellac}. The Euclidean (Matsubara) propagator
is an analytic continuation of the real time correlator
\be
G(\tau) &=& D^>(-i\tau)\\ &=& \int_{-\infty}^{\infty} d\omega
D^>(\omega) e^{-\tau\omega }\, .\ee Here we insert (\ref{Dom}) and
obtain
\be
G(\tau) &=& \int_{-\infty}^{0} d\omega
\left[1+n(\omega)\right]\sigma(\omega) e^{-\tau\omega }  +
\int_{0}^{\infty} d\omega \left[1+n(\omega)\right]\sigma(\omega)
e^{-\tau\omega }\, .\ee Now we perform a change of variables
$\omega\rightarrow -\omega~$ in the first term, and then use the
identity $1+n(\omega)+n(-\omega)=0~$ together with the property of
the spectral function, that this is an odd function in the
frequency $\sigma(-\omega)=-\sigma(\omega)~$ (the last relation is
obtained from $D^>(-\omega) = D^<(\omega)~$, which itself is
derived from the periodicity in imaginary time, known as the
Kubo-Martin-Schwinger relation \cite{lebellac}). We obtain
\be
G(\tau) &=& \int_{0}^{\infty} d\omega n(\omega)\sigma(\omega)
e^{\tau\omega }  + \int_{0}^{\infty} d\omega
\left[1+n(\omega)\right]\sigma(\omega) e^{-\tau\omega }\nonumber
\\ &=& \int_{0}^{\infty} d\omega n(\omega)\sigma(\omega)
\left[e^{\tau\omega} + e^{\beta\omega-\tau\omega}\right]\nonumber
\\ &=& \int_{0}^{\infty} d\omega \sigma(\omega) \frac{1}{e^{\beta\omega/2}
- e^{-\beta\omega/2}}\left[e^{(\tau\omega-\beta\omega/2)} +
e^{-(\tau\omega-\beta\omega/2)}\right] \, .\label{u} \ee Between
the first and the second line above we inserted the identity
$1+n(\omega)=e^{\beta\omega}n(\omega)$. From (\ref{u}) the final
result for the Matsubara correlator is obtained:
\be
G(\tau)&=& \int_{0}^{\infty} d\omega
\sigma(\omega)\frac{\cosh[\omega(\tau-\beta/2)]}{\sinh[\beta\omega/2]}
\, . \ee

%%%%%%%%%%%%%%%%%%%%%%%%%%%%%%%%
\section{Bound State Amplitudes}
\label{app-bound}

Here we derive the different relationships between the decay
constant of a given mesonic channel and the radial wave function,
for the pseudoscalar and vector channels, or its derivative, for
the scalar and axial vector channels (relations (\ref{ps-v}) and
(\ref{s-a}) in the main text).

The general idea is the following: We start from the definition of
the correlator for the mesonic channel $H$,
\be
G_H(\tau) &=& \langle  0|j_H(\tau)j_H^\dagger(0)|0\rangle\, , \ee
where $\tau$ is the Euclidean time, and the $j_H = \bar{q}
\Gamma_H q$ are quark currents bilinear in the heavy quark field
operator, that distinguishes between the scalar, vector,
pseudoscalar and axial vector channels via $\Gamma_H = 1,
\gamma_\mu, \gamma_5,$ and $\gamma_\mu\gamma_5$, respectively.
Using the Foldi-Wouthuysen-Tani transformation we decouple the
heavy quark and antiquark fields, reducing in this way the
relativistic theory of Dirac spinors to a non-relativistic theory
of Pauli spinors. Accordingly,
\be
q &=& \exp\left(\frac{\mbox{\boldmath
$\gamma$}\cdot\mbox{\boldmath $D$}}{2m}\right)\left(\ba{c} \psi\\
\chi\ea \right) =\left[1 + \frac{\mbox{\boldmath
$\gamma$}\cdot\mbox{\boldmath $D$}}{2m}+
{\cal{O}}(1/m^2)\right]\left(\ba{c} \psi\\ \chi\ea\right)\nonumber
\\ &=& \left(\ba{c} \psi\\ \chi\ea\right) +
\frac{i}{2m}\left(\ba{c} -\mbox{\boldmath
$\sigma$}\cdot\mbox{\boldmath$\stackrel{\rightarrow}{D}$}\chi\\
\mbox{\boldmath
$\sigma$}\cdot\mbox{\boldmath$\stackrel{\rightarrow}{D}$}\psi\ea\right)
+ {\cal{O}}(1/m^2) \, , \ee and
\be
{\bar{q}} = \left(\ba{c} \psi^\dagger ~~ -\chi^\dagger\ea\right) -
\frac{i}{2m}\left(\ba{c} \chi^\dagger \mbox{\boldmath
$\sigma$}\cdot\mbox{\boldmath$\stackrel{\leftarrow}{D}$} ~~~
\psi^\dagger\mbox{\boldmath
$\sigma$}\cdot\mbox{\boldmath$\stackrel{\leftarrow}{D}$}\ea\right)
+ {\cal{O}}(1/m^2)\, . \ee Here $\psi$ is the quark field that
annihilates a heavy quark, and $\chi$ the antiquark field that
creates a heavy antiquark,
\be
\psi|0\rangle=0 \, , \quad \langle 0|\psi^\dagger =0 \, , \quad
\chi^\dagger|0\rangle=0 \, , \quad \langle 0|\chi=0 \,
,\label{zero}\ee {\boldmath{$D$}} is the covariant derivative
operator, the arrows indicate its direction of action. The
Euclidean $\gamma$-matrices are given in the Dirac basis
\[ \gamma_0=\left(
\begin{array}{cc} 1 & 0 \\ 0 & -1 \end{array} \right)\, ,\qquad
\gamma_5=\left( \begin{array}{cc} 0 & 1 \\ 1 & 0
\end{array} \right)\, ,\qquad \gamma_i=\left( \begin{array}{cc} 0
& -i\sigma_i \\ i\sigma_i & 0
\end{array} \right),\]
$\sigma_i$ being the Pauli matrices. Note, that for simplicity we
suppressed the color and spin indexes on the quark field
operators. We can rewrite the meson correlator in the form
\be
G_H(\tau) = \langle  0|O_H(\tau)O_H^\dagger(0)|0\rangle\, , \ee
where $O_H$ is an operator bilinear in the quark/antiquark fields.
Its particular form will be discussed in the details that follow.
Now using
\be
O_H(\tau)=e^{H\tau}O_H(0)e^{-H\tau}\, , \ee and with the insertion
of a complete set of states $\sum_n|n\rangle\langle n|=1$, where
$|n\rangle$ is an eigenstate of the Hamiltonian,
$H|n\rangle=E_n|n\rangle$, we derive a spectral representation for
the correlator of the following form:
\be
G_H(\tau) = \sum_n |\langle  0|O_H|n\rangle|^2 e^{-E_n \tau} \,
.\ee For large times, $\tau\rightarrow\infty$, the correlator is
dominated by the state of lowest energy, and we can write
\be
G_H(\tau)\simeq f_{H} e^{-E \tau}\, , \ee where $f_H$ and $E$ are
the transition amplitude and the energy of the ground state. The
amplitude is related to the decay constant, and can be directly
related to the radial wave function, or its derivative, in the
origin, as it was shown in \cite{Bodwin:1994jh}. Following the
steps described above let us now discuss the different channels
separately.
\begin{itemize}
%%%%%
\item We start with the pseudoscalar channel, since the calculation is
simplest in this case. The current is
\be
j_{PS} &=& \bar{q}\gamma_5 q \\ &=&
\psi^\dagger\chi-\chi^\dagger\psi + {\cal{O}}(1/m)\, , \ee and to
leading order in the inverse mass the correlator is
\be
G_{PS}(\tau) &=& \langle
0|(\chi^\dagger\psi)_\tau(\psi^\dagger\chi)_{\tau=0}|0\rangle
\nonumber\\ &=& \sum_n  |\langle  0|\chi^\dagger\psi|n\rangle|^2
e^{-E_n\tau}\, .\ee At large $\tau$ the ground state dominates,
and so
\be
G_{\eta_q}(\tau)= |\langle  0|\chi^\dagger\psi|\eta_q\rangle|^2
e^{-E_{\eta_q}\tau}\, ,\ee where the $q$ refers to either of the
$c$ or $b$ quarks. The relation to the wave function in the origin
is provided by the equation (3.12a) from \cite{Bodwin:1994jh}:
\be
R_{\eta_q}(0)\equiv \sqrt{\frac{2\pi}{N_c}}\langle
0|\chi^\dagger\psi|\eta_q\rangle \, .\ee Therefore, we can easily
identify that
\be
f_{\eta_q} = \frac{N_c}{2\pi}|R_{\eta_q}(0)|^2 \, .\ee

%%%%%%
\item The scalar current is
\be
j_S &=& \bar{q} q \nonumber\\ &=& \psi^\dagger\psi -
\chi^\dagger\chi - \frac{i}{2m_q}\left[\psi^\dagger
\mbox{\boldmath$\sigma$}\cdot
\mbox{\boldmath$\stackrel{\rightarrow}{D}$}\chi +
\chi^\dagger\mbox{\boldmath$\sigma$}\cdot
\mbox{\boldmath$\stackrel{\rightarrow}{D}$}\psi +
\chi^\dagger\mbox{\boldmath$\sigma$}\cdot
\mbox{\boldmath$\stackrel{\leftarrow}{D}$}\psi +
\psi^\dagger\mbox{\boldmath$\sigma$}\cdot
\mbox{\boldmath$\stackrel{\leftarrow}{D}$}\chi \right] +
{\cal{O}}(1/m_q^2)\, .\ee  After a straightforward, but tedious
calculation, applying the relations of (\ref{zero}), and
neglecting the disconnected piece, we get the following expression
for the scalar correlator, to leading order in the inverse mass,
\be
G_S(\tau) = - \frac{1}{(2m_q)^2}\sum_n |\langle  0|\chi^\dagger
\mbox{\boldmath$\sigma$}\cdot
\mbox{\boldmath$\stackrel{\leftrightarrow}{D}$}
\psi|n\rangle|^2e^{-E_n \tau} \, .\label{chi}\ee Here the
{\boldmath$\stackrel{\leftrightarrow}{D}$} denotes the difference
of the derivative acting on the spinor to the right and on the
spinor to the left, $\chi^\dagger
\mbox{\boldmath$\stackrel{\leftrightarrow}{D}$}
\psi=\chi^\dagger\mbox{\boldmath$\stackrel{\rightarrow}{D}$}\psi -
\chi^\dagger\mbox{\boldmath$\stackrel{\leftarrow}{D}$}\psi$. For
large times the dominant contribution to the sum in (\ref{chi}) is
from the lowest lying state, the $\chi_{q0}$. Utilizing equation
(3.19b) from \cite{Bodwin:1994jh},
\be
\sqrt{\frac{3N_c}{2\pi}}R'_{\chi_{q0}}(0)\equiv
\frac{1}{\sqrt{3}}\langle
0|\chi^\dagger(\frac{1}{2}\mbox{\boldmath
$\stackrel{\leftrightarrow}{D}$}\cdot\mbox{\boldmath
$\sigma$})\psi|\chi_{q0}\rangle \, ,\ee the relation between the
amplitude and the derivative of the radial wave function at the
origin is obtained for the $\chi_{q0}$, with $q=c,b$,
\be
f_{\chi_{q0}} = -\frac{9N_c}{2\pi m^2}|R'_{\chi_{q0}}(0)|^2 \,
.\ee The $m$ refers to the corresponding $c$ or $b$ quark mass.

%%%%%
\item
Consider now the third component of the vector current:
\be
j_{V} &=& \bar{q}\gamma_i q \nonumber\\&=&
-(\psi^\dagger\sigma_i\chi + \chi^\dagger\sigma_i\psi)\, . \ee
When inserting this into the correlator this yields for the
diagonal components of this the following spectral decomposition:
\be
G_{V}(\tau) &=& \langle
0|j_{V}(\tau)j_{V}^\dagger(0)|0\rangle\nonumber\\ &=& \sum_n
|\langle  0|\chi^\dagger\sigma_i\psi|n\rangle|^2e^{-E_n \tau} \,
.\ee For $\tau\rightarrow\infty$ the leading contribution to this
correlator comes from the vector meson $V=J/\psi,\Upsilon$, and
thus
\be
G_{J/\psi}(\tau) = \langle  0|\chi^\dagger\sigma_i\psi
|V(\epsilon_i) \rangle e^{-M_{V} \tau}\, .\ee Equation (3.12b)
from \cite{Bodwin:1994jh}, with {\boldmath $\epsilon$} being the
polarization vector of the vector meson, is
\be
R_{V}(0)\mbox{\boldmath $\epsilon$}\equiv
\sqrt{\frac{2\pi}{N_c}}\langle
0|\chi^\dagger\vec{\sigma}\psi|V(\mbox{\boldmath
$\epsilon$})\rangle \, .\ee Accordingly, the transition amplitude
is
\be
\langle  0|\chi^\dagger\sigma_i\psi |V(\epsilon_i) \rangle =
\sqrt{\frac{N_c}{2\pi}}R_{V}(0)\epsilon_i \, ,\ee and summing over
all polarizations $\sum_i\epsilon_i=3$ results in
\be
f_{V} = \frac{3N_c}{2\pi}|R_{V}(0)|^2 \, .\ee

%%%%%
\item Finally, we turn to the axial vector channel. The ith component
of the current is
\be
j_{AV} &=& \bar{q}\gamma_i\gamma_5 q \nonumber\\ &=&
-\psi^\dagger\sigma_i\psi - \chi^\dagger\sigma_i\chi \nonumber \\
& & + \frac{i}{2m_q}\left[\psi^\dagger\sigma_i
\mbox{\boldmath$\sigma$}\cdot
\mbox{\boldmath$\stackrel{\rightarrow}{D}$}\chi -
\chi^\dagger\sigma_i\mbox{\boldmath$\sigma$}\cdot
\mbox{\boldmath$\stackrel{\rightarrow}{D}$}\psi +
\chi^\dagger\mbox{\boldmath$\sigma$}\cdot
\mbox{\boldmath$\stackrel{\leftarrow}{D}$}\sigma_i\psi -
\psi^\dagger\mbox{\boldmath$\sigma$}\cdot
\mbox{\boldmath$\stackrel{\leftarrow}{D}$}\sigma_i\chi \right] +
{\cal{O}}(1/m_q^2)\, .\ee When evaluating the correlator we again
make use of the cancellations of (\ref{zero}), and the
anticommutativity of the quark field operators. Furthermore, we
also apply the relation between the non-commuting Pauli matrices,
$\sigma_i\sigma_j=\delta_{ij}+i\epsilon_{ijk}\sigma_k$, and the
tensor notation of a vector product, $(\mbox{\boldmath$a$}\times
\mbox{\boldmath$b$})_i=\epsilon_{ijk}a_jb_k$. The resulting form
of the correlator, when considering only $\chi_{q1}$, with
$q=c,b$, and neglecting the disconnected part, the grounds state,
is
\be
G_{\chi_{q1}}(\tau) = - \frac{1}{(2m_q)^2}|\langle  0|\chi^\dagger
(-i\mbox{\boldmath$\stackrel{\leftrightarrow}{D}$}
\times\mbox{\boldmath$\sigma$})
\psi|\chi_{q1}(\mbox{\boldmath$\epsilon$})\rangle|^2e^{-E_{\chi_{q1}}
\tau} \, .\ee Equation (3.19c) from \cite{Bodwin:1994jh} relates
the transition amplitude to the derivative of the wave function at
the origin:
\be
\sqrt{\frac{3N_c}{2\pi}}R'_{\chi_{q1}}(0)\mbox{\boldmath
$\epsilon$}\equiv \frac{1}{\sqrt{2}}\langle
0|\chi^\dagger(-\frac{i}{2}\mbox{\boldmath
$\stackrel{\leftrightarrow}{D}$}\times\mbox{\boldmath
$\sigma$})\psi|\chi_{q1}(\mbox{\boldmath $\epsilon$})\rangle \, .
\ee We identify then the decay constant of an unpolarized axial
vector channel to be
\be
f_{\chi_{q1}} = -\frac{9N_c}{\pi m_q^2}|R'_{\chi_{q1}}(0)|^2 \,
.\ee

\end{itemize}

%%%%%%%%%%%%%%%%%%%%%%%%%%%%%%%%%%%%%%%%%%%%%
\section{Vector correlator at one-loop level}
\label{transport}

Here we derive the meson spectral function as obtained from the
imaginary part of the retarded correlator. We will focus on the
vector channel, since the spatial and the temporal component of
its spectral function contain terms relevant for the analysis of
the transport properties. The spectral function in the scalar,
pseudoscalar and axial channels can be derived in an analogous
manner.

The correlation function of a meson can be evaluated in momentum
space, using the quark propagator and its spectral representation.
Accordingly,
\be
\chi_H(\bar{\omega},\vec{p}) = N_cT\sum_n\int\frac{d^3k}{(2\pi)^3}
\mbox{Tr}\left[\Gamma_HS_F(k_0,\vec{k})\Gamma_H^\dagger
S_F^\dagger(k_0',\vec{k'})\right]\, ,\label{cor1} \ee where $N_c$
is the number of colors, and $H$ refers to a given mesonic
channel, defined by the operator
$\Gamma_H=1,\gamma_\mu,\gamma_5,\gamma_\mu\gamma_5$ for the
scalar, pseudoscalar, vector, and axial-vector channels,
respectively. Here $k_0'=k_0-\bar{\omega}$,
$\vec{k'}=\vec{k}-\vec{p}$ and $k_0=i\omega_n=i(2n+1)\pi T$. The
quark propagator has the following form \cite{lebellac}
\be
S_F(k_0,\vec{k}) = -(\gamma_0 k_0 -
\vec{\gamma}\cdot\vec{k}+m)\int_0^{1/T}d\tau
e^{k_0\tau}\int_{-\infty}^{\infty}d\omega\rho_F(\omega)
\left[1-n(\omega)\right]e^{-\omega\tau}\, . \label{qprop} \ee Here
$\rho_F(\omega)=\delta(\omega^2-\omega_k^2)$ is the free quark
spectral function, with $\omega_k^2=\vec{k}^2+m^2$ for a quark of
mass $m$. After inserting (\ref{qprop}) into (\ref{cor1}) and
evaluating the integrals over the imaginary time $\tau$, one
obtains
\be
\chi_H(\bar{\omega},\vec{p}) &=&
N_cT\sum_n\int\frac{d^3k}{(2\pi)^3}\frac{1}{\omega_k^2-k_0^2}
\frac{1}{\omega_{k'}^2-k_0'^2} \mbox{Tr}\left[\Gamma_H (\gamma^\mu
k^\mu +m)\Gamma_H^\dagger(\gamma^\nu k'^\nu +m)\right]\nonumber\\
&=& N_cT\sum_n\int\frac{d^3k}{(2\pi)^3}\frac{1}{\omega_k^2-k_0^2}
\frac{1}{\omega_{k'}^2-k_0'^2}4\left[k^\mu k'^\nu + k^\nu k'^\mu
+g^{\mu\nu}(m^2-k\cdot k') \right]\, , \label{corr2}\ee where for
the second line we specified the vector correlator,
$\Gamma_H=\gamma_\mu$.

First, let us evaluate the spatial part of (\ref{corr2}), i.e
consider $\mu,\nu=i,j$ and sum over the components:
\be
\sum_{i=j}\chi_{ij}(\bar{\omega},\vec{p}) &=&
N_c4T\sum_n\int\frac{d^3k}{(2\pi)^3} \frac{1}{\omega_k^2-k_0^2}
\frac{1}{\omega_{k'}^2-k_0'^2}\sum_{i=j}\left[k_ik_j' + k_jk_i' -
\delta_{ij}\left(m^2 -(k_0k_0'-\vec{k}
\cdot\vec{k'})\right)\right]\nonumber \\ &=& N_c
4T\sum_n\int\frac{d^3k}{(2\pi)^3} \frac{1}{\omega_k^2-k_0^2}
\frac{1}{\omega_{k'}^2-k_0'^2}\left[2k^2 + pk\cos{\theta} +
3(\omega_{k}^2-k_0^2) - 3k_0\bar{\omega}\right]\, . \ee Here we
inserted $\vec{p}\cdot\vec{k}=pk\cos{\theta}$. Using standard
techniques \cite{lebellac} the Matsubara sums can be evaluated:
\be
T\sum_n\frac{1}{\omega_k^2-k_0^2} &=&
-\frac{1}{2\omega_k}(1-2n_k)\, , \nonumber\\
T\sum_n\frac{1}{(\omega_k^2-k_0^2) (\omega_{k'}^2-k_0'^2)} &=&
-\frac{1}{4\omega_k\omega_{k'}} \left[(1-n_k-n_{k'})
\left(\frac{1}{\bar{\omega}-\omega_k-\omega_{k'}} -
\frac{1}{\bar{\omega}+\omega_k+\omega_{k'}}\right) \right.
\nonumber\\ && \left. - (n_k-n_{k'})\left(
\frac{1}{\bar{\omega}+\omega_k-\omega_{k'}} -
\frac{1}{\bar{\omega}-\omega_k+\omega_{k'}}\right) \right]\,
,\nonumber\\
T\sum_n\frac{k_0}{(\omega_k^2-k_0^2)(\omega_{k'}^2-k_0'^2)} &=&
-\frac{1}{4\omega_k} \left[(1-n_k-n_{k'})
\left(\frac{1}{\bar{\omega}-\omega_k-\omega_{k'}} +
\frac{1}{\bar{\omega}+\omega_k+\omega_{k'}}\right) \right.
\nonumber\\ && \left. + (n_k-n_{k'})\left(
\frac{1}{\bar{\omega}+\omega_k-\omega_{k'}} +
\frac{1}{\bar{\omega}-\omega_k+\omega_{k'}}\right) \right] \, ,\ee
where $n_k=1/(\exp{(\omega_k/T)}+1)$ is the Fermi distribution
function, with $\omega_k=\sqrt{k^2+m^2}$. In the following we
introduce a small momentum approximations $p\ll k$, according to
which
\be
n_{k'}-n_k &\simeq& - \frac{dn_k}{d\omega_k}
\frac{kp}{\omega_k}\cos{\theta}\, ,\\ n_{k'}+n_k &\simeq& 2n_k\,
,\\ \omega_{k'}-\omega_k&\simeq&
-\frac{pk}{\omega_k}\cos{\theta}\, ,\\
\omega_{k'}+\omega_k&\simeq& 2\omega_k \, . \ee The spectral
function is given by the imaginary part of the momentum space
retarded correlation function, $\chi_{ij}^R(\bar{\omega},\vec{p})
= \chi_{ij}(\bar{\omega}+i\epsilon,\vec{p})$,
\be
\sum_i\sigma_{ii}(\bar{\omega},\vec{p}) = -\frac{1}{\pi}
\mbox{Im}\sum_{i=j}\chi_{ij}^R(\bar{\omega},\vec{p})\, .\ee We
apply the relation \be\mbox{Im}\frac{1}{\omega-E+i\epsilon}=
-\pi\delta(\omega-E)\, ,\\ \ee and obtain
\be
\sum_i\sigma_{ii}(\bar{\omega},\vec{p}\simeq 0) &=&
N_c\int\frac{d^3k}{(2\pi)^3} \left\{\left(-\frac{2k^2
+pk\cos{\theta}}{\omega_k^2}+
\frac{3\bar{\omega}}{\omega_k}\right)(1-2n_k)\delta(\bar{\omega}
-2\omega_k) \right. \nonumber\\  & &  \qquad\qquad\qquad \left. -
\frac{2k^2+pk\cos{\theta}}{\omega_k^2}
\frac{dn}{d\omega_k}\frac{pk}{\omega_k}\cos{\theta}\left[
\delta\left(\bar{\omega}-\frac{pk}{\omega_k}\cos{\theta}\right) -
\delta\left(\bar{\omega}+\frac{pk}{\omega_k}\cos{\theta}\right)\right]
\right. \nonumber\\  & &  \qquad\qquad\qquad \left. +
\frac{3\bar{\omega}}{\omega_k}
\frac{dn}{d\omega_k}\frac{pk}{\omega_k}\cos{\theta} \left[
\delta\left(\bar{\omega}-\frac{pk}{\omega_k}\cos{\theta}\right) +
\delta\left(\bar{\omega}+\frac{pk}{\omega_k}\cos{\theta}\right)\right]
\right\} \\  &=&
N_c\int\frac{d^3k}{(2\pi)^3}\left[\left(-\frac{2k^2+pk\cos{\theta}}{\omega_k^2}+
\frac{3\bar{\omega}}{\omega_k}\right)(1-2n_k)\delta(\bar{\omega}
-2\omega_k) \right. \nonumber\\  & &  \qquad\qquad\qquad \left. -2
\frac{2k^2}{\omega_k^2}
\frac{dn}{d\omega_k}\frac{pk}{\omega_k}\cos{\theta}
\delta\left(\bar{\omega}-\frac{pk}{\omega_k}\cos{\theta}\right)\right]
\, .\ee
In the integrand $x\delta(\omega-x) \rightarrow
\omega\delta(\omega)$  for the limit of zero momentum,
$x\rightarrow 0$. The result for the spectral function at zero
momentum is then
\be
\sum_i\sigma_{ii}(\bar{\omega},\vec{p}=0) &=& \frac{N_c}{8\pi^2}
\bar{\omega}^2\sqrt{1-\frac{4m^2}{\bar{\omega}^2}}\tanh{\frac{\bar{\omega}}{4T}}
\left(2+\frac{4m^2}{\bar{\omega}^2}\right) - 2N_c
\bar{\omega}\delta(\bar{\omega})\int\frac{d^3k}{(2\pi)^3}
\frac{dn}{d\omega_k}\frac{2k^2}{\omega_k^2} \, .\label{sii}\ee The
first term is the leading order perturbative result for the
continuum, also obtained in \cite{Karsch:2003wy}. The second term
provides a temperature-dependent contribution to the correlator.
We evaluate this term in a non-relativistic approximation at the
end of this Appendix.

Next, we evaluate the temporal part of the correlator. Inserting
$\mu=\nu=0$ in (\ref{corr2}) results in
\be
\chi_{00}(\bar{\omega},\vec{p}) &=&
N_c4T\sum_n\int\frac{d^3k}{(2\pi)^3} \frac{1}{\omega_k^2-k_0^2}
\frac{1}{\omega_{k'}^2-k_0'^2}\left[2k_0k_0' + \left(m^2
-(k_0k_0'-\vec{k} \cdot\vec{k'})\right)\right]\nonumber \\ &=&
N_c4T\sum_n\int\frac{d^3k}{(2\pi)^3} \frac{1}{\omega_k^2-k_0^2}
\frac{1}{\omega_{k'}^2-k_0'^2}\left[2\omega_k^2-pk\cos{\theta} -
(\omega_{k}^2-k_0^2) - k_0\bar{\omega}\right]\, .\ee The
corresponding component of the spectral function at small momentum
is
\be
\sigma_{00}(\bar{\omega},\vec{p}\simeq 0) &=& -\frac{1}{\pi}
\mbox{Im}\chi_{00}(\bar{\omega},\vec{p}\simeq 0)\\ &=&
N_c\int\frac{d^3k}{(2\pi)^3}\left[\left(-\frac{2\omega_k^2-pk\cos{\theta}}{\omega_k^2}
+ \frac{\bar{\omega}}{\omega_k}\right)(1-2n_k)\delta(\bar{\omega}
-2\omega_k) \right. \nonumber \\ & & \qquad\qquad\qquad \left. + 4
\frac{dn}{d\omega_k}\frac{pk}{\omega_k}\cos{\theta}\delta\left(\bar{\omega}
-\frac{pk}{\omega_k}\cos{\theta}\right)\right]\, . \ee In the
limit of zero momentum the contribution from the first term
vanishes, and the final result is
\be
\sigma_{00}(\bar{\omega},\vec{p}=0) &=& 4N_c
\bar{\omega}\delta(\bar{\omega})\int\frac{d^3k}{(2\pi)^3}
\frac{dn_k}{d\omega_k} \\ &=& -\chi_s
\bar{\omega}\delta(\bar{\omega})\, .\ee  Here we identified the
static charge susceptibility, $\chi_s$, which can be evaluated
explicitly. For heavy quarks the Fermi distribution function is
well approximated by the Boltzmann distribution,
$n_k=\exp{(-\omega_k/T)}$, and thus
\be
\chi_s &=& -4N_c\int\frac{d^3k}{(2\pi)^3}
\frac{dn_k}{d\omega_k}\nonumber\\ &=& 4N_c\frac{1}{2\pi^2 T}
\int_0^\infty dk k^2 e^{-\frac{\omega_k}{T}} \nonumber\\ &=&
4N_c\frac{1}{2\pi^2 T} \int_{m}^\infty d\omega \omega
(\omega^2-m^2)^{1/2}e^{-\frac{\omega_k}{T}}\\ &=&
4N_c\frac{1}{2\pi^2} m^2 K_2\left(\frac{m}{T}\right)\, , \ee where
$K_2$ is the modified Bessel function of the second kind. Since
$m/T\gg1$ the following approximation can be applied
\be
K_\nu(z)\rightarrow \sqrt{\frac{\pi}{2z}}e^{-z}
\quad{\mbox{for}}\quad z\gg 1\, , \ee yielding for the static
susceptibility
\be
\chi_s = 4N_c\frac{1}{(2\pi)^{3/2}} m^{3/2} T^{1/2} e^{-m/T}\, .
\label{suscept}\ee

Finally, we now evaluate the second contribution to (\ref{sii}),
while incorporating a non-relativistic approximation for the heavy
quarks. This means $n_k=\exp{(-\omega_k/T)}$ and $k/m\ll 1$. Thus
\be
\sum_i\sigma_{ii}(\bar{\omega},\vec{p}=0) &=& -4N_c
\bar{\omega}\delta(\bar{\omega})\int\frac{d^3k}{(2\pi)^3}
\frac{dn}{d\omega_k}\frac{k^2}{\omega_k^2}\nonumber\\ &=& 4N_c
\bar{\omega}\delta(\bar{\omega})\frac{1}{2\pi^2 T}
e^{-m/T}\int_0^\infty dk
\frac{k^4}{k^2+m^2}e^{-k^2/(2mT)}\nonumber\\ &\simeq&
\frac{4N_c}{2\pi^2 Tm^2}\bar{\omega}\delta(\bar{\omega})
e^{-m/T}\int_0^\infty dk k^4e^{-k^2/(2mT)}\nonumber\\ &=&
3\bar{\omega}\delta(\bar{\omega}) \left(4N_c\frac{1}{(2\pi)^{3/2}}
m^{3/2} T^{1/2} e^{-m/T}\right)\frac{T}{m}\nonumber\\ &=& 3\chi_s
\frac{T}{m}\bar{\omega}\delta(\bar{\omega})  \, .\ee

%%%%%%%%%%%%%%%%%%%%%%%%%%%%%%%%%%%%%
\section{Numerical results as tables}
\label{app:tables}

Here we provide the results of the numerical calculations for the
properties of the charmonia states, in Table \ref{tab:c}, and
bottomonia states, in Table \ref{tab:b}, together with the
temperature dependence of the screening mass in the screened
Cornell potential, as well as the parameters of the potential
fitted to the lattice internal energy and the charmonia properties
in Tables \ref{tab:latc} and \ref{tab:latb}.

\begin{table}[h]\nonumber
\begin{minipage}{10cm}
\begin{ruledtabular}
\caption{Results for the $c\bar{c}$ mesons.}
\begin{tabular}{||c|c|c|c|c|c||}
$T/T_c$& $\mu$[GeV]& state& M[GeV]& $\langle
r^2\rangle^{1/2}$[fm]&$|R(0)|^2$(S); $|R'(0)|^2$(P) \\ \hline  0&
0& 1S& 3.07& 0.453& 0.735 GeV$^3$\\ & &2S&3.67&0.875&0.534
GeV$^3$\\ & & 1P& 3.5& 0.695& 0.06 GeV$^5$\\ \hline 1.1& 0.25& 1S&
3.044& 0.574& 0.468 GeV$^3$\\ & &2S&3.338&1.591&0.137 GeV$^3$\\ &
& 1P& 3.294& 1.124& 0.012 GeV$^5$\\ \hline 1.25& 0.314& 1S& 3.031&
0.633& 0.395 GeV$^3$\\ & & 1P& 3.23& 1.58& 0.005 GeV$^5$\\ \hline
1.5& 0.395& 1S& 3.012& 0.73& 0.311 GeV$^3$\\ & & 1P& 3.132& 1.776&
0.001 GeV$^5$\\ \hline 1.75& 0.472& 1S& 2.992& 0.89& 0.23
GeV$^3$\\ \hline  2& 0.55& 1S& 2.968& 1.2& 0.15 GeV$^3$ \nonumber
\end{tabular}
\label{tab:c}
\end{ruledtabular}
\end{minipage}
\end{table}

\begin{table}[h]
\begin{minipage}{10cm}
\begin{ruledtabular}
\caption{Results for the $b\bar{b}$ mesons.}
\begin{tabular}{||c|c|c|c|c|c||}
$T/T_c$&$\mu$[GeV]&state&M[GeV]& $\langle
r^2\rangle^{1/2}$[fm]&$|R(0)|^2$(S); $|R'(0)|^2$(P)\\ \hline
0&0&1S&9.445&0.225&9.422 GeV$^3$\\ & &2S&10.004&0.508&4.377
GeV$^3$\\ & &1P&9.897&0.407&1.39 GeV$^5$\\ & &3S&10.355&0.74&3.42
GeV$^3$\\ & & 2P&10.259&0.65&1.74 GeV$^5$\\ \hline
1.1&0.25&1S&9.52&0.243&8.346 GeV$^3$\\ & &2S&9.945&0.641&2.632
GeV$^3$\\ & &1P&9.893&0.501&0.738 GeV$^5$\\ & &2P&10.092&0.95&0.54
GeV$^5$\\ \hline 1.25&0.314&1S&9.536&0.249&8.07 GeV$^3$\\ &
&2S&9.924&0.695&2.195 GeV$^3$\\ & &1P&9.885&0.541&0.574 GeV$^5$\\
& &2P&10.041&1.13&0.32 GeV$^5$\\ \hline
1.5&0.395&1S&9.553&0.259&7.644 GeV$^3$\\ & &2S&9.895&0.806&1.676
GeV$^3$\\ & &1P&9.872&0.603&0.404 GeV$^5$\\ \hline
1.75&0.472&1S&9.569&0.267&7.257 GeV$^3$\\ & &2S&9.864&0.974&1.096
GeV$^3$\\ & &1P&9.854&0.715&0.255  GeV$^5$\\ \hline
2&0.55&1S&9.582&0.277&6.818 GeV$^3$\\ & &2S&9.832&1.246&0.627
GeV$^3$\\ & &1P&9.832&1.053&0.12 GeV$^5$\\ \hline
2.25&0.628&1S&9.594&0.289&6.369 GeV$^3$\\  & &1P&9.802&2.404&0.015
GeV$^5$\\ \hline 2.5&0.705&1S&9.603&0.309&5.867 GeV$^3$\\ \hline
3&0.86&1S&9.618&0.342&4.923 GeV$^3$\\ \hline
3.5&1.015&1S&9.627&0.41&3.865 GeV$^3$\\ \hline
4&1.17&1S&9.63&0.545&2.771 GeV$^3$\\ \hline
4.5&1.325&1S&9.628&0.78&1.707 GeV$^3$\\ \hline
5&1.48&1S&9.62&2.146&0.578 GeV$^3$
\end{tabular}
\label{tab:b}
\end{ruledtabular}
\end{minipage}
\end{table}

\begin{table}[h]
\begin{minipage}{12cm}
\begin{ruledtabular}
\caption{Results for the $c\bar{c}$ mesons using the potential
fitted to lattice internal energy.}
\begin{tabular}{||c|c|c|c|c|c|c|c||}
$T/T_c$&$\sigma$[GeV]$^2$&$\mu$[GeV]&$C$[GeV]&state&M[GeV]&$\langle
r^2\rangle^{1/2}$[fm]&$|R(0)|^2$(S); $|R'(0)|^2$(P)\\ \hline
0&0.18&0&0&1S&3.177&0.500&0.441 GeV$^3$\\ & & & &
2S&3.733&0.923&0.370 GeV$^3$\\ & & & & 1P& 3.534&0.733&0.034
GeV$^5$
\\ \hline 1.07&0.0417&0.123&1.467&1S&3.522&0.399&0.863 GeV$^3$\\ &
& & &1P& 4.059&0.791&0.072 GeV$^5$\\ \hline
1.13&0.0484&0.155&1.09&1S&3.448&0.458&0.688 GeV$^3$ \\ \hline
1.25&0.074&0.193&0.747&1S&3.448&0.636&0.411 GeV$^3$\\
%\hline 1.40&0&0.313&0.611&1S&3.236&1.072&0.217 GeV$^3$
\end{tabular}
\label{tab:latc}
\end{ruledtabular}
\end{minipage}
\end{table}

\begin{table}[h]
\begin{minipage}{12cm}
\begin{ruledtabular}
\caption{Results for the $b\bar{b}$ mesons using the potential
fitted to lattice internal energy.}
\begin{tabular}{||c|c|c|c|c|c|c|c||}
$T/T_c$&$\sigma$[GeV]$^2$&$\mu$[GeV]&$C$[GeV]&state&M[GeV]&$\langle
r^2\rangle^{1/2}$[fm]&$|R(0)|^2$(S); $|R'(0)|^2$(P)\\ \hline
0&0.181&0&0&1S&9.693&0.283&3.477 GeV$^3$\\ & & & &
2S&10.115&0.567&2.268 GeV$^3$ \\ & & & & 3S&10.419&0.797&1.949
GeV$^3$ \\ & & & & 1P& 9.996&0.452&0.539 GeV$^5$ \\ & & & & 2P&
10.315&0.701&0.773 GeV$^5$ \\ \hline
1.07&0.074&0.131&1.471&1S&9.839&0.227&5.590 GeV$^3$ \\ & & & &
2S&10.544&0.454&3.932 GeV$^3$ \\ & & & & 3S&10.930&0.927&1.508
GeV$^3$ \\ & & & &1P& 10.318&0.346&1.775 GeV$^5$\\ & & & &2P&
10.826&0.630&1.742 GeV$^5$\\ \hline
1.13&0.087&0.161&1.094&1S&9.827&0.234&5.267 GeV$^3$ \\ & & & &
2S&10.435&0.530&2.829 GeV$^3$ \\ & & & &1P& 10.266&0.377&1.369
GeV$^5$\\ \hline 1.25&0.134&0.174&0.752&1S&9.826&0.236&5.136
GeV$^3$\\  & & & & 2S&10.435&0.530&2.771 GeV$^3$ \\ & & &
&1P&10.266&0.376&1.475 GeV$^5$\\ \hline
1.40&0&0.408&0.608&1S&9.820&0.258&4.695 GeV$^3$\\ \hline
1.95&0&0.313&0.611&1S&9.864&0.388&3.216 GeV$^3$
\end{tabular}
\label{tab:latb}
\end{ruledtabular}
\end{minipage}
\end{table}

%%%%%%%%%%%%%%%%%%

\end{document}